\documentclass[11pt,a4paper]{article}

\usepackage[margin=1in]{geometry}
\usepackage{lscape}
\usepackage{multirow}
\usepackage{amsmath}
\usepackage{amsthm}
\allowdisplaybreaks
\usepackage{amsfonts}
\usepackage[normalem]{ulem}
\usepackage{amssymb}
\usepackage{bbm}
\usepackage{extarrows}
\usepackage{float}
\usepackage{bm}
\usepackage{tikz}
\usepackage[caption=false]{subfig}
\usepackage{graphicx}
\usepackage{array}
\usepackage[pagebackref,linktoc=all]{hyperref}
\hypersetup{colorlinks,linkcolor=red,citecolor=blue,urlcolor=lightgray}
\usepackage[capitalise]{cleveref}
\usepackage[inline]{enumitem}

\makeatletter
\def\uwave{\bgroup \markoverwith{\lower4.5\p@\hbox{\sixly \textcolor{blue}{\char58}}}\ULon}
\font\sixly=lasy6 
\def\uwavered{\bgroup \markoverwith{\lower4.5\p@\hbox{\sixly \textcolor{red}{\char58}}}\ULon}
\font\sixly=lasy6 
\makeatother

\title{Learning Reserve Prices in Second-Price Auctions}

\author{
Yaonan Jin\thanks{Columbia University. Email: {\tt yj2552@columbia.edu}}
\and
Pinyan Lu\thanks{Shanghai University of Finance and Economics. Email: {\tt lu.pinyan@mail.shufe.edu.cn}}
\and
Tao Xiao\thanks{Huawei TCS Lab. Email: {\tt xiaotao21@huawei.com}}
}

\date{}

\usepackage{algpseudocode}
\usepackage{algorithm}

\theoremstyle{plain}
\newtheorem{theorem}{Theorem}

\newtheorem{lemma}{Lemma}[section]

\newtheorem{fact}[lemma]{Fact}
\newtheorem{corollary}[lemma]{Corollary}

\theoremstyle{definition}

\theoremstyle{remark}

\newtheoremstyle{restate}{}{}{\itshape}{}{\bfseries}{~(restated).}{.5em}{\thmnote{#3}}
\theoremstyle{restate}

\crefname{theorem}{Theorem}{Theorems}
\crefname{problem}{Problem}{Problems}
\crefname{openproblem}{Open Problem}{Open Problems}
\crefname{lemma}{Lemma}{Lemmas}
\crefname{claim}{Claim}{Claims}
\crefname{proposition}{Proposition}{Propositions}
\crefname{fact}{Fact}{Facts}
\crefname{corollary}{Corollary}{Corollaries}
\crefname{conjecture}{Conjecture}{Conjectures}
\crefname{assumption}{Assumption}{Assumptions}

\crefname{definition}{Definition}{Definitions}
\crefname{condition}{Condition}{Conditions}
\crefname{example}{Example}{Examples}

\crefname{remark}{Remark}{Remarks}

\crefname{equation}{Equation}{Equations}
\crefname{figure}{Figure}{Figures}

\newcommand{\E}{\operatorname{{\bf E}}}

\renewcommand{\Pr}{\operatorname{{\bf Pr}}}

\newcommand{\Var}{\operatorname{{\bf Var}}}

\newcommand{\supp}{\mathbf{supp}}

\renewcommand{\d}{\mathrm{d}}    
\newcommand{\lhs}{\mathrm{LHS}} 
\newcommand{\rhs}{\mathrm{RHS}} 
\newcommand{\indicator}{\mathbbm{1}}
\newcommand{\dominate}{\succeq}
\newcommand{\bedominated}{\preceq}

\newcommand{\argmax}{\mathop{\rm argmax}}
\newcommand{\eqdef}{\stackrel{\textrm{\tiny def}}{=}}
\newcommand{\eps}{\varepsilon}

\newcommand{\vect}[1]{\boldsymbol{#1}}
\newcommand{\RR}{\mathbb{R}}
\newcommand{\RRP}{\RR_{\geq 0}}
\newcommand{\NN}{\mathbb{N}}
\newcommand{\NNP}{\NN_{\geq 1}}

\newcommand{\val}{v}
\newcommand{\bval}{\overline{val}}

\newcommand{\sample}{\vect{s}}
\newcommand{\samples}{\vect{S}}
\newcommand{\samplei}[1][j]{s_{#1}}
\newcommand{\osamplei}[1][j]{\widehat{s}_{#1}}
\newcommand{\sampleti}[2]{s_{#1,~#2}}
\newcommand{\hsamplei}[1][i]{\widehat{\vect{s}}_{#1}}
\newcommand{\hsampleti}[2]{\widehat{s}_{#1,~#2}}
\newcommand{\bsamplet}[1][t]{\overline{\mathrm v}_{#1}}

\newcommand{\distri}[1][j]{F_{#1}}
\newcommand{\jdistr}{\mathcal{F}}
\newcommand{\jdistri}[1][i]{\jdistr_{#1}}

\newcommand{\edistr}{\mathcal{E}}
\newcommand{\edistri}[1][i]{\edistr_{#1}}

\newcommand{\sdistr}{\widetilde{\jdistr}}
\newcommand{\sdistri}[1][i]{\sdistr_{#1}}

\newcommand{\tdistr}{\jdistr^*}
\newcommand{\stdistr}{\widetilde{\jdistr}^*}
\newcommand{\tdistri}[1][i]{\jdistr_{#1}^*}
\newcommand{\stdistri}[1][i]{\widetilde{\jdistr}_{#1}^*}

\newcommand{\sedistr}{\widetilde{\edistr}}
\newcommand{\sedistri}[1][i]{\sedistr_{#1}}

\newcommand{\shade}{{\sf S}_{\mathcal{F}}}
\newcommand{\eshade}{{\sf S}_{\mathcal{E}}}

\newcommand{\ar}{{\sf AR}}

\newcommand{\arm}{\sf Anonymous Reserve}
\newcommand{\optm}{\sf Myerson Auction}

\newcommand{\cO}{\mathcal{O}}

\newcommand{\cOmega}{\Omega}
\newcommand{\tO}{\widetilde{\mathcal{O}}}
\newcommand{\tTheta}{\widetilde{\Theta}}
\newcommand{\tOmega}{\widetilde{\Omega}}

\newcommand{\mhr}{\textsc{MHR}}

\begin{document}
\maketitle

\begin{abstract}
This paper proves the tight sample complexity of {\sf Second-Price Auction with Anonymous Reserve}, up to a logarithmic factor, for each of all the value distribution families studied in the literature: $[0,\, 1]$-bounded, $[1,\, H]$-bounded, regular, and monotone hazard rate (MHR).
Remarkably, the setting-specific tight sample complexity $\mathsf{poly}(\varepsilon^{-1})$ depends on the precision $\varepsilon \in (0, 1)$, but not on the number of bidders $n \geq 1$.
Further, in the two bounded-support settings, our learning algorithm allows {\em correlated} value distributions.

In contrast, the tight sample complexity $\tilde{\Theta}(n) \cdot \mathsf{poly}(\varepsilon^{-1})$ of {\sf Myerson Auction} proved by Guo, Huang and Zhang (STOC~2019) has a nearly-linear dependence on $n \geq 1$, and holds only for {\em independent} value distributions in every setting.

We follow a similar framework as the Guo-Huang-Zhang work, but replace their information theoretical arguments with a direct proof.
\end{abstract}

\setcounter{page}{0}
\thispagestyle{empty}

\newpage

\section{Introduction}
\label{sec:intro}

Bayesian auction theory assumes that the seller knows the prior value information of bidders and would design auctions/mechanisms by leveraging that information. In real-life applications, the priors are learned from historical data. How much data is needed to learn good auctions? This question motivates the research interest in the sample complexity for auction design, initiated by Cole and Roughgarden \cite{CR14}.\footnote{A very related topic, the sample complexity of optimal pricing for a single bidder, dates back to \cite{DRY15}. Also, some regret-minimization variants date earlier to \cite{BHW02,BKRW03,BH05}.} Concretely, it focuses on how many samples are needed, regarding the precision $\eps \in (0, 1)$ and the bidder population $n \in \NNP$, to learn an $(1 - \eps)$-approximate optimal auction. A long line of work had improved the sample complexity~\cite{CR14,MR15,DHP16,GN17,S17,HMR18}. The recent breakthrough result by Guo et al.\ \cite{GHZ19} derived the tight sample complexity, up to poly-logarithmic factors, for all the value distribution families considered in the literature.

The above results all target the {\em revenue-optimal} single-item auction, namely the canonical {\optm} \cite{M81}. Nonetheless, {\optm} is fairly complicated and rarely used in real life \cite{LMP06}. In contrast, the AGT community has placed ``simplicity'' as a primary goal for auction design \cite{HR09,CHMS10,AHNPY19,JLTX20,JLQTX19,JJLZ22}. In practice, one of the most popular auctions is {\sf Second-Price Auction with Anonymous Reserve} (i.e., setting the same reserve price for every bidder), e.g., the auctions in eBay, AdX, and Adsense. We emphasize that such an auction has straightforward instructions:
\begin{quote}
    {\em Based on the bidders' value distributions, the seller carefully selects a reserve $r \in \RRP$ for the item. If all the bids are below $r$, then the seller retains the item. If only the highest bid reaches $r$, then the highest bidder wins the item by paying this reserve price. Otherwise (i.e., two or more bids reach $r$), the highest bidder wins the item by paying a price of the second-highest bid.}
\end{quote}
The significance and practicality of {\arm} naturally motivate a rich literature to study its approximability against {\optm} in terms of revenues \cite{HR09,H13,AHNPY19,JLTX20,JLQTX19,JLQ19} and learnability \cite{CGM15,MR15,MM16,RS16,HLW18}.

The first ``approximability'' result was attained in Myerson's original paper \cite{M81}: When the value distributions are i.i.d.\ and satisfy the standard {\em regularity} assumption (see \Cref{subsec:prelim:distr} for its definition), {\optm} reduces to {\arm}. Even if the distributional assumptions are greatly relaxed, as we quote from \cite{HR09}: ``{\em In quite general settings, simple auctions like {\arm} provably approximates the optimal expected revenue, to within a small constant factor.}'' Moreover, its learnability has been tackled in various contexts. For example, Cesa-Bianchi et al.\ \cite{CGM15} assumed i.i.d.\ and $[0, 1]$-bounded value distributions and got a nearly optimal sample complexity of $\tTheta(\eps^{-2})$.\footnote{Precisely, Cesa-Bianchi et al.\ studied the slightly different problem of regret minimization over a time horizon $t \in \mathrm{T}$ (in terms of the cumulative revenue loss against the optimal {\arm}). While imposing a strong distributional assumption, the seller is assumed to know just the allocations and payments in the past rounds. Cesa-Bianchi et al.\ obtained a nearly optimal $\tTheta(\sqrt{\mathrm{T}})$-regret algorithm. This regret bound easily indicates the $\tTheta(\eps^{-2})$ sample complexity bound.}

\begin{table}[t]
\centering
{\small}
\begin{tabular}{|c||>{\centering\arraybackslash}p{2.3cm}|>{\centering\arraybackslash}p{3.1cm}||>{\centering\arraybackslash}p{2.3cm}|>{\centering\arraybackslash}p{1.55cm}|>{\centering\arraybackslash}p{1.55cm}|}
\hline
\rule{0pt}{12pt} & \multicolumn{2}{c||}{{\optm}} & \multicolumn{3}{c|}{{\arm}} \\ [1pt]
\hline
\hline
\rule{0pt}{12pt}$[0, 1]$-additive & $\tTheta(n \cdot \eps^{-2})$ & \multirow{4}*{\rule{0pt}{20pt}\cite{GHZ19}} & $\tTheta(\eps^{-2})$ & & \\ [1pt]
\cline{1-2}\cline{4-4}
\rule{0pt}{12pt}$[1, H]$ & $\tTheta(n \cdot H \cdot \eps^{-2})$ & & $\tTheta(H \cdot \eps^{-2})$ & \cite{HMR18} & \multirow{2}*{\rule{0pt}{12pt}Thm~\ref{thm:upper}} \\ [1pt]
\cline{1-2}\cline{4-4}
\rule{0pt}{12pt}regular & $\tTheta(n \cdot \eps^{-3})$ & & $\tTheta(\eps^{-3})$ & & \\ [1pt]
\cline{1-2}\cline{4-5}
\rule{0pt}{12pt}$\mhr$ & $\tTheta(n \cdot \eps^{-2})$ & & $\tTheta(\eps^{-2})$ & \cite{GHZ19} & \\ [1pt]
\hline
\end{tabular}
\caption{For {\optm}, the nearly-tight bounds in all settings are obtained by \cite{GHZ19}. For {\arm}, the upper bounds in all settings follow from our \Cref{thm:upper}, and the matching lower bounds are proved by \cite{HMR18,GHZ19}. (In the $\mhr$ setting, the above lower bounds hold for {\em discrete} $\mhr$ distributions, but the best-known lower bounds for {\em continuous} $\mhr$ distributions are just $\tOmega(n \cdot \eps^{-3/2})$ and $\tOmega(\eps^{-3/2})$ \cite{HMR18,GHZ19}.)}
\label{tbl:result}
\end{table}


Despite the above discussions, how many samples do we need to learn an $(1 - \eps)$-approximate optimal reserve price when the bidders have (possibly) {\em distinct} value distributions? This problem is essential to understand {\arm} but remains unsettled. Because we only need to learn a good reserve price, conceivably, the task should be much easier than learning {\optm} (which requires a complete understanding of all bidders' distributions). Our work shows that this is precisely the case: As \Cref{tbl:result} illustrates, {\arm} in comparison has dramatically smaller sample complexity. Remarkably, it depends only on the precision $\eps \in (0, 1)$ but not on the population $n \in \NNP$.

Our learning algorithm for {\arm} is clear and intuitive and thus may be more attractive in practice. First, we slightly ``shrink'' (in the sense of stochastic dominance) the empirical distributions determined by the samples, resulting in the {\em dominated empirical distributions}. Then, we compute the optimal reserve price for these dominated empirical distributions (or, when there are multiple optimal reserve prices, any of them).\footnote{This reserve price must be {\em bounded} since the (dominated) empirical distributions determined by the samples are {\em bounded} almost surely (even in the {\em unbounded} regular/$\mhr$ settings).} Employing this reserve price turns out to generate an $(1 - \eps)$-fraction as much revenue as the optimal {\arm}.

This framework was proposed by \cite{GHZ19}, and the analysis has two parts: {\em revenue monotonicity} and {\em revenue smoothness}. The revenue monotonicity of a specific auction means if a distribution instance $\jdistr$ stochastically dominates another $\jdistr'$, then the two revenues satisfy that $\mbox{\sc Rev}(\jdistr) \geq \mbox{\sc Rev}(\jdistr')$. Since {\optm} and {\arm} both have this feature, for the analysis of revenue monotonicity, we can apply arguments \`{a} la \cite{GHZ19}. Moreover, revenue smoothness means if two distribution instances are stochastically close (in some metric), then their revenues must also be close. Guo et al.\ establish the revenue smoothness of {\optm} via an elegant {\em information theoretical argument}. However, this proof scheme is inapplicable here, and instead, we will present a more direct proof.

Before elaborating on the new argument, let us briefly explain why {\arm} needs much fewer samples. The outcome of such an auction (i.e., the allocation and the payment) relies on the highest and second-highest bids, whose distributions suffice to determine the optimal reserve price. (In contrast, we must know the distributions of all bidders to implement {\optm}.) Since only two distributions rather than $n$ distributions are involved, we can eliminate the dependence of the sample complexity on the population.

Nonetheless, the restriction on the highest and second-highest bids incurs another issue. In the model, we assume the bids to be {\em mutually independent}. This assumption is critical for the information theoretical arguments by Guo et al.\ and the optimality of {\optm}. Conversely, the highest two bids, in general, are {\em correlated}. It is highly non-trivial whether we can extend the information theoretical arguments to accommodate the correlated distributions.\footnote{The information theoretical arguments by \cite{GHZ19} crucially rely on a particular form of {\em Pinsker's inequality}, which holds only for {\em independent} distributions. For {\arm}, in contrast, we need to deal with the generally {\em correlated} highest and second-highest distributions. So, we must abandon the proof scheme by Guo et al.\ and directly reason about {\arm} revenues.} Thus, we prove the revenue smoothness by working directly with {\arm} revenue. The techniques derived here may find more applications in the future. (For example, they complement the extreme value theorems by \cite{CR14,CD15,MR15}.) We believe that a similar approach, associated with the tools by \cite{JLQTX19}, can circumvent the information theoretical arguments by \cite{GHZ19} and refine the poly-logarithmic factors in their sample complexity of {\optm}.

\vspace{.1in}
\noindent
{\bf Correlation.}
Another benefit of the direct arguments is that even if the bids are {\em arbitrarily correlated} (but capped with a specific high value), learning {\arm} needs the same amount of samples. This generalized model is arguably much more realistic. In this direction, an intriguing open problem is to study, given the correlated distributions, the sample complexity of the optimal mechanisms \cite{DFK15,PP15} or the optima in certain families of {\em robust} mechanisms \cite{R01,CHLW11,BGLT19}.

\vspace{.1in}
\noindent
{\bf Data Compression.}
If we care about the {\em space complexity} of the learning algorithms, the improvement on {\arm} against {\optm} is even more significant. To learn {\optm}, we need $\tO(n^2) \cdot \mathsf{poly}(1 / \eps)$ space both to implement the algorithm and to store the output auction. (Note that each sample is an $n$-dimensional value vector.) Namely, we cannot predict the future bids and must record all details of the learned ``virtual value functions''. However, for {\arm}, since only the highest and second-highest bids are involved, we only need $\mathsf{poly}(1 / \eps)$ space to implement the algorithm and $\cO(1)$ space to store the learned reserve price. This property is crucial to large markets, where historical data cannot be stored entirely in the memory, and we wish to handle it in very few passes (in the sense of streaming algorithms).

\subsection{Comparison with Previous Approaches}

To understand the sample complexity of {\arm}, an immediate attempt is to readopt the algorithm of \cite{CGM15}, under minor modification to accommodate non-identical and even correlated value distributions (rather than just the i.i.d.\ ones). However, that algorithm crucially relies on a particular property of i.i.d.\ value distributions: we can infer $\jdistri[1]$ point-wise from $\jdistri[2]$ and vice versa, where $\jdistri$ denotes the CDF of the $i$-th highest bid. Without the i.i.d.\ assumption, the correlation between $\jdistri[1]$ and $\jdistri[2]$ is much more complex, which makes this attempt fail to work for our purpose.

Also, one may attempt the empirical revenue maximization scheme, which gives the nearly tight sample complexity for the similar task ``optimal pricing $p_{j} \eqdef \argmax_{p} p \cdot (1 - F_{j}(p))$ for a single bidder $F_{j}$''. However, in the regular and the $\mhr$ settings, the proof of either sample complexity crucially relies on the underlying distributional assumption \cite{DRY15,HMR18}.
For example, given a regular/$\mhr$ distribution $F_{j}$, either the optimal price $p_{j}$ is unique, or all the optimal prices $p_{j}$ form a connected interval.
In contrast, given $n \in \NNP$ regular/$\mhr$ value distributions, $\jdistri[1]$ and $\jdistri[2]$ can have $\Omega(n)$ disconnected optimal prices $p_{i} \eqdef \argmax_{p} p \cdot (1 - \jdistri(p))$ (see \cite[Example~2]{JLTX20}).
Accordingly, $\jdistri[1]$ and $\jdistri[2]$ themselves cannot be regular/$\mhr$, which rejects this attempt as well.

Another approach in the literature is to construct an $\eps$-net of all candidate reserve prices, namely a $\mathsf{poly}(1 / \eps)$-size hypothesis set $\mathcal{H}$, and figure out the best one in $\mathcal{H}$ through the samples (see \cite{DHP16,RS16,GN17}, which use this method to learn {\optm}). In fact, for the $[0, 1]$-bounded and $[1, H]$-bounded settings, it is a folklore that $\eps$-net type algorithms can attain the (nearly) tight sample complexity. However, the regular and $\mhr$ settings are less understood due to the lack of suitable tools, such as some particular extreme value theorems. Here we address this question; given the developed techniques, we {\em can} present such sample-optimal $\eps$-net type algorithms in both settings.

However, we prefer the ``shrink-then-optimize'' framework of \cite{GHZ19} for two reasons. First, $\eps$-net type algorithms choose distinct hypothesis sets $\mathcal{H}$ for different value distribution families, i.e., the distributional assumption somehow is part of the ``input''. By contrast, the new framework gives a {\em unified} and robust learning algorithm. In particular, different distributional assumptions induce different sample complexities but do not affect the algorithm implementation. Second, our paper demonstrates that the new framework works not only for the input value distributions as in \cite{GHZ19} but also for some ``sketched'' distributions, i.e., order statistics, for our purpose. It would be interesting to see further extensions of this framework.

\subsection{Other Related Work}

As mentioned, after the pioneering work of \cite{CR14}, the sample complexity of {\optm} had been improved in a sequence of papers \cite{MR15,DHP16,RS16,GN17,S17} and was finally answered by \cite{GHZ19}. En route, many techniques have been developed and may be helpful to mechanism design, learning theory, and information theory. For an outline of these techniques, the reader can turn to \cite[Section~1]{GHZ19}.

Another related topic is the sample complexity of {\em single-bidder} revenue maximization. Now, the optimal mechanism is to post the {\em monopoly price} $\mathrm{p} \eqdef \argmax \big\{\val \cdot \big(1 - \distri[](\val)\big): \val \in \RRP\big\}$ and then let the bidder make a {\em take-it-or-leave-it} decision. Again, the problem is self-contained only under one of the four assumptions in \Cref{tbl:result}. Up to a poly-logarithmic factor, the optimal sample complexity is $\tTheta(\eps^{-2})$ in the $[0, 1]$-bounded additive-error setting \cite{BBHM08,HMR18}, $\tTheta(H \cdot \eps^{-2})$ in the $[1, H]$-bounded setting \cite{BBHM08,HMR18}, $\tTheta(\eps^{-3})$ in the continuous regular setting \cite{DRY15,HMR18}, and $\tTheta(\eps^{-2})$ in the $\mhr$ setting \cite{GHZ19}.

One can easily see that,\footnote{E.g., imagine there is a dominant bidder in revenue maximization, and the other $(n - 1)$ bidders are negligible.} in each of the four settings, the sample complexity of {\arm} must be lower bounded by the single-bidder sample complexity. Since each mentioned single-bidder lower bound matches with the claimed sample complexity of {\arm} in \Cref{tbl:result} (up to a logarithmic factor), it remains to establish the upper bounds in the bulk of this work.

To learn good posted prices for a single buyer, a complementary direction is to investigate how much expected revenue is achievable using exactly {\em one sample}. When the distribution is regular, \cite{DRY15} showed that using the sampled value as the price guarantees half of the optimal revenue. Indeed, this ratio is the best possible (in the sense of worst-case analysis) when the seller must post a {\em deterministic} price.
However, better ratios are possible under certain adjustments to the model. First, if the seller can access the {\em second} sample, he can improve the ratio to $0.509$ \cite{BGMM18}.\footnote{Concretely, \cite{BGMM18} employs the empirical revenue maximization pricing scheme. That is, let $\samplei[1] \geq \samplei[2]$ be the two samples, then choose $\samplei[1]$ as the posted price when $\samplei[1] \geq 2 \cdot \samplei[2]$ and choose $\samplei[2]$ otherwise.} Second, if a {\em randomized} price is allowed, the seller can get a better revenue guarantee by constructing a 
particular price distribution from the single sample \cite{FILS15}. Recently, \cite{ABB22} improved this ratio to $0.501$, and proved that no randomized pricing scheme could achieve a $0.511$-approximation.
Moreover, if the buyer's distribution satisfies the stronger $\mhr$ condition, \cite{HMR18} gave a deterministic $0.589$-approximation one-sample pricing scheme. Afterward, \cite{ABB22} improved this ratio to $0.644$, and obtained a $0.648$ impossibility result for any deterministic/randomized pricing scheme.

Another motivation of the ``mechanism design via sampling'' program is the recent research interest in {\em multi-item} mechanism design, where {\optm} or its naive generalizations are no longer optimal.
The optimal multi-item mechanisms are often computationally/conceptually hard \cite{DDT14,CDOPSY22,CDPSY18,CMPY18}. Instead, a rich literature proves that {\em simple} multi-item mechanisms are learnable from polynomial samples and constantly approximate the optimal revenues \cite{MR15,BSV16,DS22,MR16,CD17,S17,BSV18,GW21,GHTZ21}.

\vspace{.1in}
\noindent
{\bf Organization.}
Notation and preliminaries are given below. In \Cref{sec:sample}, we show our learning algorithm (see \Cref{alg:main}) and present the analysis of revenue monotonicity. In \Cref{sec:revenue}, we present the analysis of revenue smoothness, hence the sample complexity promised in \Cref{tbl:result}. In \Cref{sec:conclusion}, we conclude this paper with a discussion on future research directions.

\section{Notation and Preliminaries}
\label{sec:prelim}

{\bf Notation.}
Denote by $\RRP$ (resp.\ $\NNP$) the set of all non-negative real numbers (resp.\ positive integers). For any pair of integers $b \geq a \geq 0$, denote by $[a]$ the set $\{1, 2, \cdots, a\}$, and by $[a: b]$ the set $\{a, a + 1, \cdots, b\}$. Denote by $\indicator(\cdot)$ the indicator function. The function $(\cdot)_+$ maps a real number $z \in \RR$ to $\max\{0, z\}$. For convenience, we interchange bid/value and bidder/buyer.

\subsection{Probability}
\label{subsec:prelim:distr}

We use the calligraphic letter $\jdistr$ to denote an {\em input instance} (i.e., an $n$-dimensional {\em joint distribution}), from which the buyers $j \in [n]$ draw a value vector $\sample = (\samplei)_{j \in [n]} \in \RRP^n$. Particularly, if the value $\samplei$'s are {\em independent} random variables (drawn from a {\em product distribution} $\jdistr$), we further write $\jdistr = \{\distri\}_{j \in [n]}$, where each $\distri$ presents the marginal value distribution of the individual buyer $j \in [n]$. Regarding the {\arm} auctions (to be elaborated in \Cref{subsec:prelim:mech}), the highest and second-highest values $\osamplei[1]$ and $\osamplei[2]$ are of particular interest. We respectively denote by $\jdistri[1]$ and $\jdistri[2]$ the distributions of $\osamplei[1]$ and $\osamplei[2]$.

As usual in the literature, we use the notations $\jdistr$ and $\jdistri$ (for $i \in \{1, 2\}$) and $\distri$ (for $j \in [n]$) also to denote the corresponding CDF's. However, we assume that a single-dimensional CDF $\jdistri$ or $\distri$ is left-continuous,\footnote{For the $n$-dimensional input distribution $\jdistr$, we never work with its CDF directly.} in the sense that if a buyer has a random value $s \sim F$ for a price-$p$ item, then his {\em unwilling-to-purchase probability} is $\Pr[s < p]$ rather than $\Pr[s \leq p]$. Further, we say a distribution $F$ {\em stochastically dominates} another distribution $F'$ (or simply $F \dominate F'$) when their CDF's satisfy $F(\val) \leq F'(\val)$ for any $\val \in \RRP$.

We investigate the input instance $\jdistr$ in four canonical settings. The first and second settings, where the support $\supp(\jdistr)$ is bounded within the $n$-dimensional hypercube $[0, 1]^n$ or $[1, H]^n$ (for a given real number $H \geq 1$), are clear.

In the third setting, the input instance is a product distribution $\jdistr = \{\distri\}_{j \in [n]}$, where each $\distri$ is a {\em continuous regular distribution}.\footnote{More precisely, $\distri$ can have a unique probability mass at its support supremum.} Denote by $f_j$ the corresponding PDF. According to \cite{M81}, the regularity means the {\em virtual value function}
\[
    \varphi_j(\val) \eqdef \val - \tfrac{1 - \distri(\val)}{f_j(\val)}
\]
is monotone non-decreasing on the support $\supp(\distri)$.

In the last setting, the input instance $\jdistr = \{\distri\}_{j \in [n]}$ is also a product distribution, but each $\distri$ now may be a {\em discrete} or {\em continuous} (or even {\em mixture}) distribution that has a {\em monotone hazard rate} ($\mhr$). Let us specify the $\mhr$ condition \cite{BMP63} in the next paragraph.

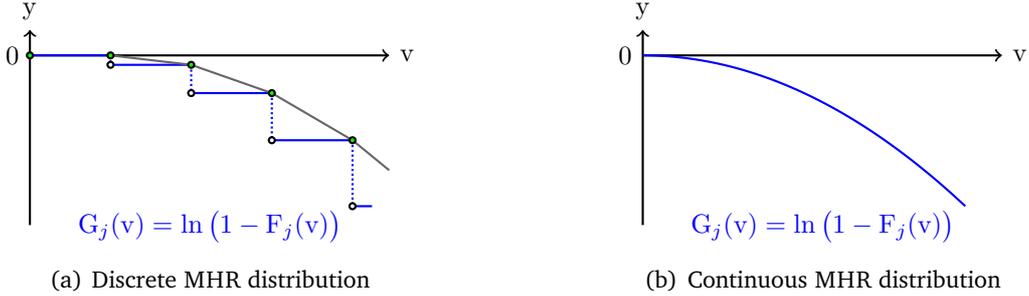
\begin{figure}[htbp]
\centering
\subfloat[Discrete $\mhr$ distribution
\label{fig:prelim:mhr:1}]{
\begin{tikzpicture}[thick, smooth, scale = 2.25]
\draw[->] (0, 0) -- (2.1, 0);
\draw[->] (0, -1) -- (0, 0.15);
\node[above] at (0, 0.15) {\small $\mathrm{y}$};
\node[right] at (2.1, 0) {\small $\val$};
\node[left] at (0, 0) {\small $0$};
\node[anchor = north east, color = blue] at (1.8856, -0.85) {\small $\mathrm{G}_j(\val) = \ln\big(1 - \distri(\val)\big)$};
\draw[color = blue] (0, 0) -- (0.4714, 0) (0.4714, -0.0556) -- (0.9428, -0.0556) (0.9428, -0.2222) -- (1.4142, -0.2222) (1.4142, -0.5) -- (1.8856, -0.5) (1.8856, -0.8888) -- (2, -0.8888);
\draw[densely dotted, color = blue] (0.4714, 0) -- (0.4714, -0.0556) (0.9428, -0.0556) -- (0.9428, -0.2222) (1.4142, -0.2222) -- (1.4142, -0.5) (1.8856, -0.5) -- (1.8856, -0.8888);
\draw[color = darkgray!80] (0.4714, 0) -- (0.9428, -0.0556) -- (1.4142, -0.2222) -- (1.8856, -0.5) -- (2.1, -0.6769);
\draw[color = black, fill = green] (0.4714, 0) circle(0.5pt);
\draw[color = black, fill = green] (0.9428, -0.0556) circle(0.5pt);
\draw[color = black, fill = green] (1.4142, -0.2222) circle(0.5pt);
\draw[color = black, fill = green] (1.8856, -0.5) circle(0.5pt);
\draw[color = black, fill = white] (0.4714, -0.0556) circle(0.5pt);
\draw[color = black, fill = white] (0.9428, -0.2222) circle(0.5pt);
\draw[color = black, fill = white] (1.4142, -0.5) circle(0.5pt);
\draw[color = black, fill = white] (1.8856, -0.8888) circle(0.5pt);
\draw[color = black, fill = green] (0, 0) circle(0.5pt);
\end{tikzpicture}}
\hspace{2cm}
\subfloat[Continuous $\mhr$ distribution
\label{fig:prelim:mhr:2}]{
\begin{tikzpicture}[thick, smooth, scale = 2.25]
\draw[->] (0, 0) -- (2.1, 0);
\draw[->] (0, -1) -- (0, 0.15);
\node[above] at (0, 0.15) {\small $\mathrm{y}$};
\node[right] at (2.1, 0) {\small $\val$};
\node[left] at (0, 0) {\small $0$};
\node[anchor = north east, color = blue] at (1.8856, -0.85) {\small $\mathrm{G}_j(\val) = \ln\big(1 - \distri(\val)\big)$};
\draw[color = blue, domain = 0: 1.8856] plot (\x, {-\x^2 / 4});
\end{tikzpicture}}
\caption{Demonstration for discrete and continuous $\mhr$ distributions.}
\label{fig:prelim:mhr}
\end{figure}

\vspace{.1in}
\noindent
{\bf $\mhr$ Distribution.}
A {\em discrete} $\mhr$ instance $\jdistr = \{\distri\}_{j \in [n]}$ must be supported on a discrete set $\{k \Delta: k \in \NNP\}$ (as \Cref{fig:prelim:mhr:1} demonstrates), where $\Delta > 0$ is a given step-size. For each $j \in [n]$, consider the {\em step function} $\mathrm{G}_j(\val) \eqdef \ln\big(1 - \distri(\val)\big)$ (marked in blue) as well as the {\em piece-wise linear function} $\mathrm{L}_j$ (marked in gray) determined by the origin $(0,~0)$ and the {\em ``\raisebox{-1ex}{\LARGE $\urcorner$}''-type points} $\big(k \cdot \Delta,~\mathrm{G}_j(k \cdot \Delta)\big)$'s (marked in green). The $\mhr$ condition holds iff each $\mathrm{L}_j$ is a concave function. Moreover, for a {\em continuous} $\mhr$ instance $\jdistr = \{\distri\}_{j \in [n]}$, each individual $\distri$ is supported on a possibly distinct interval. The $\mhr$ condition holds iff each $\mathrm{G}_j(\val) \eqdef \ln\big(1 - \distri(\val)\big)$ is a concave function on its own support, as \Cref{fig:prelim:mhr:2} illustrates.

\vspace{-.1in}

\subsection{Anonymous Reserve}
\label{subsec:prelim:mech}

In a {\sf Second-Price Auction with {\arm}}, the seller posts an {\em a priori} reserve $r \in \RRP$ to the item. There are three possible outcomes: (i)~when no buyer has a value of at least the reserve $r$, the auction would abort; (ii)~when there is exactly one such buyer, he would pay the reserve $r$ for winning the item; (iii)~when there are two or more such buyers, the highest-value buyer (with arbitrary tie-breaking rule) would pay the second-highest value (i.e., a price of at least the reserve $r$) for winning the item.

We now formulate the {\em expected revenue} from the above mechanism \cite[Fact~1]{CGM15}. Sample a random value vector $\sample = (\samplei)_{j \in [n]} \sim \jdistr$ and then denote by $(\osamplei[1],~\osamplei[2])$ the highest and second-highest values. By simulating the mechanism, we have
\begin{align*}
\mbox{(outcome revenue)}
& = r \cdot \indicator(\osamplei[1] \geq r > \osamplei[2]) + \osamplei[2] \cdot \indicator(\osamplei[2] \geq r) \\
& = r \cdot \indicator(\osamplei[1] \geq r) + (\osamplei[2] - r)_+ \\
\Rightarrow \qquad
\mbox{(expected revenue)}
& = r \cdot \Pr[\osamplei[1] \geq r] + \E[(\osamplei[2] - r)_+].
\end{align*}
In order to comprehend the expected revenue (denoted by $\ar(r,~\jdistr)$ for brevity), we need to know nothing (e.g.\ the correlation between $\osamplei[1]$ and $\osamplei[2]$) but the {\em marginal} CDF's $\jdistri[1]$ and $\jdistri[2]$. So, we may write $\jdistr = \jdistri[1] \uplus \jdistri[2]$, namely the ``union'' of the highest and second-highest CDF's. Equipped with the new notations, let us formulate the expected revenue more explicitly.

\begin{fact}[Revenue Formula {\cite{CGM15}}]
\label{fact:ar_revenue}
Under any reserve $r \in \RRP$, the corresponding {\arm} auction extracts an expected revenue of
\[
\mbox{$\ar(r,~\jdistr) = r \cdot \big(1 - \jdistri[1](r)\big) + \int_{r}^{\infty} \big(1 - \jdistri[2](x)\big) \cdot \d x$}.
\]
\end{fact}

When the reserve $r \in \RRP$ is selected optimally, namely $r_{\jdistr} \eqdef \argmax \{\ar(r,~\jdistr): r \in \RRP\}$ (which might be infinity), we simply write $\ar(\jdistr) = \ar(r_{\jdistr},~\jdistr)$. Based on the revenue formula in \Cref{fact:ar_revenue}, one can easily check the next \Cref{fact:revenue_monotonicity} via elementary algebra.

\begin{fact}
\label{fact:revenue_monotonicity}
The following holds for any pair of instances $\jdistr = \jdistri[1] \uplus \jdistri[2]$ and $\jdistr' = \jdistri[1]' \uplus \jdistri[2]'$ that admits the stochastic dominance $\jdistri[1] \dominate \jdistri[1]'$ and $\jdistri[2] \dominate \jdistri[2]'$:
\begin{enumerate}[font = {\em \bfseries}]
\item $\ar(r,~\jdistr) \geq \ar(r,~\jdistr')$ for any reserve $r \in \RRP$.
\item $\ar(\jdistr) \geq \ar(\jdistr')$.
\end{enumerate}
\end{fact}

For ease of presentation, we also need the extra notations below, and the next \Cref{fact:shade_function} (see \Cref{subapp:fact:shade_function} for its proof) will often be invoked in our later proof.
\begin{itemize}
    \item The parameter $\beta \eqdef \frac{\ln(8m / \delta)}{m}$, in which $m \in \NNP$ represents the sample complexity and $\delta \in (0, 1)$ denotes the {\em failing probability} of a learning algorithm.
    
    \item The empirical instance $\edistr = \edistri[1] \uplus \edistri[2]$ is given by a number of $m \in \NNP$ samples. Consider the $i$-th highest entry of every sample, then the $i$-th highest empirical distribution $\edistri$ is exactly the {\em uniform distribution} supported on these $i$-th highest entries. Equivalently, $\edistri$ is the {\em uniform distribution} induced by $m$ samples from the $i$-th highest distribution $\jdistri$.
    
    \item The shaded instance $\sdistr = \sdistri[1] \uplus \sdistri[2]$: every $i$-th highest shaded CDF $\sdistri$ is defined as $\sdistri(\val) \eqdef \shade(\jdistri(\val))$ for all value $\val \in \RRP$, where the function
    
    \hspace{2cm}
    $\shade(x) \eqdef \min \big\{1,~~x + \sqrt{8\beta \cdot x \cdot (1 - x)} + 7\beta\big\}$,
    \hspace{1cm}
    $\forall x \in [0, 1]$.
\item The shaded empirical instance $\sedistr = \sedistri[1] \uplus \sedistri[2]$: every $i$-th highest shaded empirical CDF $\sedistri$ is defined as $\sedistri(\val) \eqdef \eshade(\edistri(\val))$ for all value $\val \in \RRP$, where the function

    \hspace{2cm}
    $\eshade(x) \eqdef \min \big\{1,~~x + \sqrt{2\beta \cdot x \cdot (1 - x)} + 4\beta\big\}$,
    \hspace{1cm}
    $\forall x \in [0, 1]$.
\end{itemize}

\begin{fact}
\label{fact:shade_function}
Both of $\shade(x)$ and $\eshade(x)$ are non-decreasing functions on interval $x \in [0, 1]$.
\end{fact}

Regarding \Cref{fact:shade_function}, all the above instances are well defined. Without ambiguity, we may write $\sdistri \eqdef \shade(\jdistri)$ and $\sedistri \eqdef \eshade(\edistri)$. In the next section, we will show certain properties of/among them and the input instance $\jdistr = \jdistri[1] \uplus \jdistri[2]$.

\section{Empirical Algorithm}
\label{sec:sample}

In this section, we first present our learning algorithm and formalize our main results (given respectively in \Cref{alg:main,thm:upper}). Afterwards, we probe into the learned {\arm} auction via the {\em revenue monotonicity} (cf.\ \Cref{fact:revenue_monotonicity}). As a result, the learning problem will be converted into proving a certain property (parameterized by $\beta = \frac{\ln(8m / \delta)}{m}$, where $m$ is the sample complexity) of the concerning instance $\jdistr = \jdistri[1] \uplus \jdistri[2]$.

\begin{algorithm}[H]
\caption{Empirical Algorithm}
\textbf{Input:}
sample matrix $\samples = (\sampleti{t}{j})_{m \times n}$, where each row $(\sampleti{t}{j})_{j \in [n]}$ is a sample drawn from $\jdistr$

\textbf{Output:} an $(1 - \eps)$-approximately optimal {\arm} auction for instance $\jdistr$
\begin{algorithmic}[1]
\ForAll {$i \in \{1,~2\}$} \do
    
    \State Let $\hsamplei = (\hsampleti{t}{i})_{t \in [m]}$ be the row-wise $i$-th highest entries of the sample matrix $\samples$
    
    \Statex \hfill // Namely, reorder rows $(\sampleti{t}{j})_{j \in [n]}$ so that $\sampleti{t}{(1)} \geq \cdots \geq \sampleti{t}{(n)}$, then $\hsampleti{t}{i} \eqdef \sampleti{t}{(i)}$
    
    \State Let $\edistri$ be the $i$-th highest empirical CDF induced by the $i$-th highest sample $\hsamplei$
    
    \State Let $\sedistri \eqdef \eshade(\edistri)$ be the shaded counterpart of the $i$-th highest empirical CDF $\edistri$
\EndFor
\State \Return the optimal reserve $r_{\sedistr}$ for $\sedistr = \sedistri[1] \uplus \sedistri[2]$ (under any tie-breaking rule)
\end{algorithmic}
\label{alg:main}
\end{algorithm}

\begin{theorem}
\label{thm:upper}
With $(1 - \delta)$ confidence, the reserve $r_{\sedistr} \in \RRP$ output by \Cref{alg:main} gives a nearly optimal {\arm} revenue $\ar(r_{\sedistr},~\jdistr) \geq \ar(\jdistr) - \eps$, conditioned on
\begin{enumerate}[font = {\em \bfseries}]
\item $m = \cO\big(\eps^{-2} \cdot (\ln\eps^{-1} + \ln\delta^{-1})\big)$ and the instance $\jdistr$ is supported on $[0, 1]^{n}$.
\end{enumerate}
Alternatively, $\ar(r_{\sedistr},~\jdistr) \geq (1 - \eps) \cdot \ar(\jdistr)$, conditioned on
\begin{enumerate}[font = {\em \bfseries}]
\setcounter{enumi}{1}
\item $m = \cO\big(\eps^{-2} \cdot H \cdot (\ln\eps^{-1} + \ln H + \ln\delta^{-1})\big)$ and the instance $\jdistr$ is supported on $[1, H]^{n}$.
\item $m = \cO\big(\eps^{-3} \cdot (\ln\eps^{-1} + \ln\delta^{-1})\big)$ and the instance $\jdistr$ is regular.
\item $m = \cO\big(\eps^{-2} \cdot (\ln\eps^{-1} + \ln\delta^{-1})\big)$ and the instance $\jdistr$ is $\mhr$.
\end{enumerate}
\end{theorem}

\noindent
{\bf Analysis via Revenue Monotonicity.}
The following Lemma~\ref{lem:empirical} suggests that (with high confidence) the empirical instance $\edistr = \edistri[1] \uplus \edistri[2]$ is close to the original instance $\jdistr = \jdistri[1] \uplus \jdistri[2]$ in the Kolmogorov distance. We defer the proof of Lemma~\ref{lem:empirical} to \Cref{subapp:lem:empirical}.

\begin{lemma}
\label{lem:empirical}
With $(1 - \delta)$ confidence, for both $i \in \{1,~2\}$, the following holds for the $i$-th highest CDF $\jdistri$ and its empirical counterpart $\edistri$: for any value $\val \in \RRP$,
\begin{align*}
\big|\edistri(\val) - \jdistri(\val)\big|
\leq \sqrt{2\beta \cdot \jdistri(\val) \cdot \big(1 - \jdistri(\val)\big)} + \beta.
\end{align*}
\end{lemma}

By construction ($\sedistri \eqdef \eshade(\edistri)$ for both $i \in \{1,~2\}$), the shaded empirical instance $\sedistr = \sedistri[1] \uplus \sedistri[2]$ must be dominated by the empirical instance $\edistr = \edistri[1] \uplus \edistri[2]$, thus likely being dominated by the original instance $\jdistr = \jdistri[1] \uplus \jdistri[2]$ as well (in view of Lemma~\ref{lem:empirical}).

Instead, let us consider the shaded instance $\sdistr = \sdistri[1] \uplus \sdistri[2]$ derived directly from the original instance $\jdistr$ via the other function $\shade(\cdot)$, i.e., $\sdistri \eqdef \shade(\jdistri)$ for both $i \in \{1,~2\}$. Compared to the earlier function $\eshade(\cdot)$, the current function $\shade(\cdot)$ distorts the input $x \in [0, 1]$ to a greater extent:
\begin{align*}
\shade(x)
& = \mbox{$\min \big\{1,~~x + \sqrt{8\beta \cdot x \cdot (1 - x)} + 7\beta\big\}$} \\
& \geq \mbox{$\min \big\{1,~~x + \sqrt{2\beta \cdot x \cdot (1 - x)} + 4\beta\big\}$}
= \eshade(x),
\end{align*}
where the inequality is strict when $\shade(x) < 1$. Given these and in view of Lemma~\ref{lem:empirical} (that the empirical instance $\edistr$ is close to the original instance $\jdistr$), the two shaded instances $\sedistr$ and $\sdistr$ are likely to admit the dominance $\sedistri \dominate \sdistri$ for both $i \in \{1,~2\}$.

These two propositions are formalized as Lemma~\ref{lem:dominate:distr} (see \Cref{subapp:lem:dominate:distr} for its proof):

\begin{lemma}
\label{lem:dominate:distr}
In the case of Lemma~\ref{lem:empirical}, which happens with $(1 - \delta)$ confidence, for both $i \in \{1,~2\}$, the following holds for the empirical $i$-th highest CDF $\sedistri$:
\begin{enumerate}[font = {\em \bfseries}]
\item $\sedistri(\val) \geq \jdistri(\val)$ for any $\val \in \RRP$, i.e., $\sedistri$ is dominated by the given $i$-th highest CDF $\jdistri$.
\item $\sedistri(\val) \leq \sdistri(\val)$ for any $\val \in \RRP$, i.e., $\sedistri$ dominates the shaded $i$-th highest CDF $\sdistri$.
\end{enumerate}
\end{lemma}

Using the reserve $r_{\sedistr}$ output by \Cref{alg:main}, the corresponding {\arm} auction extracts an expected revenue of $\ar(r_{\sedistr},~\jdistr)$ from the original instance $\jdistr = \jdistri[1] \uplus \jdistri[2]$. Below, we give a lower bound of this revenue, which is more convenient for later analysis.

\begin{lemma}
\label{lem:dominate:revenue}
In the case of Lemma~\ref{lem:empirical}, which happens with $(1 - \delta)$ confidence, from the original instance $\jdistr = \jdistri[1] \uplus \jdistri[2]$, the {\arm} with a reserve of $r_{\sedistr}$ generates a revenue better than the optimal {\arm} revenue from the shaded instance $\sdistr = \sdistri[1] \uplus \sdistri[2]$:
\[
\ar(r_{\sedistr},~\jdistr)
\geq \ar(r_{\sdistr},~\sdistr)
= \ar(\sdistr).
\]
\end{lemma}

\begin{proof}
Due to Lemma~\ref{lem:dominate:distr} and \Cref{fact:revenue_monotonicity} (i.e., the revenue monotonicity with respect to {\arm}):
\begin{align*}
\ar(r_{\sedistr},~\jdistr)
& \geq \ar(r_{\sedistr},~\sedistr)
&& \mbox{\tt (Part~1 of Lemma~\ref{lem:dominate:distr}: dominance $\jdistri \dominate \sedistri$)} \\
& \geq \ar(r_{\sdistr},~\sedistr)
&& \mbox{\tt ($r_{\sedistr}$ is optimal to $\sedistr$ but $r_{\sdistr}$ may not be)} \\
& \geq \ar(r_{\sdistr},~\sdistr)
= \ar(\sdistr).
&& \mbox{\tt (Part~2 of Lemma~\ref{lem:dominate:distr}: dominance $\sedistri \dominate \sdistri$)}
\end{align*}
This completes the proof of Lemma~\ref{lem:dominate:revenue}.
\end{proof}

Remarkably, the lower-bound revenue $\ar(\sdistr)$ is irrelevant to \Cref{alg:main}, since we directly construct the shaded instance $\sdistr = \sdistri[1] \uplus \sdistri[2]$ from the original instance $\jdistr$ via the function $\shade(\cdot)$ (parameterized by $\beta = \frac{\ln(8m / \delta)}{m}$, where $m$ is the promised sample complexity). Based on the above discussions, \Cref{thm:upper} immediately follows if we have
\begin{align*}
& \ar(\sdistr) \geq \ar(\jdistr) - \eps
&& \mbox{\tt (the $[0, 1]$-bounded setting)} \\
& \ar(\sdistr) \geq (1 - \eps) \cdot \ar(\jdistr)
&& \mbox{\tt (the other three settings)}
\end{align*}
These two inequalities will be justified in \Cref{sec:revenue}.

\section{Revenue Smoothness}
\label{sec:revenue}

In this section, we will bound the additive or multiplicative revenue gap between the shaded instance $\sdistr = \sdistri[1] \uplus \sdistri[2]$ and the original instance $\jdistr = \jdistri[1] \uplus \jdistri[2]$. First of all, one can easily check the next \Cref{fact:parameter} via elementary algebra.

\begin{fact}
\label{fact:parameter}
The following holds for the parameter $\beta \eqdef \frac{\ln(8m / \delta)}{m}$:
\begin{enumerate}[font = {\em \bfseries}]
\item $\beta \leq \frac{\eps^2}{12}$ when $m \geq 36\eps^{-2} \cdot (\ln\eps^{-1} + \ln\delta^{-1} + 3)$.
    \hfill
    {\em ($[0, 1]$-bounded setting)}
\item $\beta \leq \frac{\eps^2 \cdot H^{-1}}{48}$ when $m = 144\eps^{-2} \cdot H \cdot (\ln\eps^{-1} + \ln H + \ln\delta^{-1} + 4)$.
    \hfill
    {\em ($[1, H]$-bounded setting)}
\item $\beta \leq \frac{\eps^3}{2880}$ when $m \geq 11520\eps^{-3} \cdot (\ln\eps^{-1} + \ln\delta^{-1} + 4)$.
    \hfill
    {\em (continuous regular setting)}
\item $\beta \leq \frac{\eps^2}{1870}$ when $m \geq 5610\eps^{-2} \cdot (\ln\eps^{-1} + \ln\delta^{-1} + 5)$.
    \hfill
    {\em ($\mhr$ setting)}
\end{enumerate}
\end{fact}

\subsection{\texorpdfstring{$[0, 1]$-Bounded Setting}{}}
\label{subsec:revenue:0_1_support}

Given the sample complexity $m = \cO\big(\eps^{-2} \cdot (\ln\eps^{-1} + \ln\delta^{-1})\big)$ promised in Part~1 of \Cref{thm:upper}, we safely assume $m \geq 36\eps^{-2} \cdot (\ln\eps^{-1} + \ln\delta^{-1} + 3)$. Consider the function $\shade(\cdot)$: for $x \in [0, 1]$,
\begin{align*}
\shade(x)
& = \mbox{$\min \big\{1,~~x + \sqrt{8\beta \cdot x \cdot (1 - x)} + 7\beta\big\}$} \\
& \leq \mbox{$x + \sqrt{8\beta \cdot x \cdot (1 - x)} + 7\beta$} \\
& \leq \mbox{$x + \sqrt{2\beta} + 7\beta$}
&& \mbox{\tt (as $x \cdot (1 - x) \leq \frac{1}{4}$)} \\
& \leq \mbox{$x + \frac{1}{\sqrt{6}} \cdot \eps + \frac{7}{12} \cdot \eps^2$}
&& \mbox{\tt (Part~1 of \Cref{fact:parameter}: $\beta \leq \frac{\eps^2}{12}$)} \\
& \leq \mbox{$x + \eps$},
&& \mbox{\tt (as $\frac{1}{\sqrt{6}} + \frac{7}{12} \approx 0.9916 < 1$)}
\end{align*}
which means that $\sdistri(\val) \leq \jdistri(\val) + \eps$ for all value $\val \in [0, 1]$ and both $i \in \{1,~2\}$. Let $r_{\jdistr} \in [0, 1]$ denote the optimal reserve for the original instance $\jdistr = \jdistri[1] \uplus \jdistri[2]$. Thus,\footnote{Note that the interval of integration can be safely truncated to the support supremum of $s_u = 1$.}
\begin{align*}
\ar(\jdistr) - \ar(\sdistr)
& \leq \ar(r_{\jdistr},~\jdistr) - \ar(r_{\jdistr},~\sdistr)
\qquad \mbox{\tt ($r_{\jdistr}$ may not be optimal for $\sdistr$)} \\
& = \mbox{$r_{\jdistr} \cdot \big(\sdistri[1](r_{\jdistr}) - \jdistri[1](r_{\jdistr})\big) + \int_{r_{\jdistr}}^{1} \big(\sdistri[2](x) - \jdistri[2](x)\big) \cdot \d x$} \\
& \leq \mbox{$r_{\jdistr} \cdot \eps + \int_{r_{\jdistr}}^{1} \eps \cdot \d x$}
= \eps.
\end{align*}
This concludes the proof in the setting with $[0, 1]$-bounded support.

\subsection{\texorpdfstring{$[1, H]$-Bounded Setting}{}}
\label{subsec:revenue:1_H_support}

Given the sample complexity $m = \cO\big(\eps^{-2} \cdot H \cdot (\ln\eps^{-1} + \ln H + \ln\delta^{-1})\big)$ promised in Part~2 of \Cref{thm:upper}, we safely assume $m \geq 144\eps^{-2} \cdot H \cdot (\ln\eps^{-1} + \ln H + \ln\delta^{-1} + 4)$. To see this amount of samples is sufficient to learn a nearly optimal {\arm}, the next two facts will be useful.

\begin{fact}
\label{fact:1_H:1}
From the original instance $\jdistr = \jdistri[1] \uplus \jdistri[2]$, the optimal {\arm} revenue $\ar(\jdistr)$  is at least the support infimum of $s_l = 1$.
\end{fact}

\begin{proof}
Obvious, e.g.\ the item always gets sold out under a reserve of $1$.
\end{proof}

\begin{fact}
\label{fact:1_H:2}
For the original instance $\jdistr = \jdistri[1] \uplus \jdistri[2]$, there is an optimal {\arm} auction having a reserve of $r_{\jdistr} \in \big[1,~\jdistri[1]^{-1}(\frac{H - 1}{H})\big] \subseteq [1, H]$.
\end{fact}

\begin{proof}
When there are multiple alternative optimal reserves $r_{\jdistr}$'s, we would select the {\em smallest} one. Clearly, the bound $\jdistri[1]^{-1}(\frac{H - 1}{H})$ is at least the support infimum of $s_l = 1 \leq \jdistri[1]^{-1}(0)$. Actually, employing the reserve of $1$ guarantees as much revenue as employing another reserve $r \in \big(\jdistri[1]^{-1}(\frac{H - 1}{H}),~H\big]$: recall the {\arm} revenue formula,
\begin{align*}
\ar(1,~\jdistr) - \ar(r,~\jdistr)
& = \mbox{$1 \cdot \big(1 - \jdistri[1](1)\big) - r \cdot \big(1 - \jdistri[1](r)\big) + \int_{1}^{r} \big(1 - \jdistri[2](x)\big) \cdot \d x$} \\
& \geq \mbox{$1 \cdot \big(1 - \jdistri[1](1)\big) - r \cdot \big(1 - \jdistri[1](r)\big)$}
\hspace{1cm} \mbox{\tt ($r > \jdistri[1]^{-1}(\frac{H - 1}{H}) \geq 1$)} \\
& \geq \mbox{$1 \cdot \big(1 - \jdistri[1](1)\big) - H \cdot (1 - \frac{H - 1}{H})$}
\hspace{1.12cm} \mbox{\tt ($r \leq H$ \& $\jdistri[1](r) > \frac{H - 1}{H}$)} \\
& = 1 - 1 = 0.
\hspace{4.58cm} \mbox{\tt (as $\jdistri[1](1) = 0$)}
\end{align*}
That is, under our tie-breaking rule, any reserve $r \in \big(\jdistri[1]^{-1}(\frac{H - 1}{H}),~H\big]$ cannot be optimal, which completes the proof of \Cref{fact:1_H:1}.
\end{proof}

Define a parameter $\mathrm{B} \eqdef {\jdistri[2]}^{-1}(\frac{H - 1}{H}) \in [1, H]$. As shown in the former $[0, 1]$-bounded setting, the function $\shade(x) \leq x + \sqrt{8\beta \cdot x \cdot (1 - x)} + 7\beta$ for $x \in [0, 1]$. We deduce that\footnote{Note that the interval of integration can be safely truncated to the support supremum of $s_u = H$.}
\begin{align}
\notag
\ar(\jdistr) - \ar(\sdistr)
& \leq \ar(r_{\jdistr},~\jdistr) - \ar(r_{\jdistr},~\sdistr)
\hspace{1cm} \mbox{\tt ($r_{\jdistr}$ may not be optimal to $\sdistr$)} \\
\notag
& = \mbox{$r_{\jdistr} \cdot \big(\sdistri[1](r_{\jdistr}) - \jdistri[1](r_{\jdistr})\big) + \int_{r_{\jdistr}}^{H} \big(\sdistri[2](x) - \jdistri[2](x)\big) \cdot \d x$} \\
\label{eq:1_H:0}
& \leq \mbox{\rm (First Term)} + \mbox{\rm (Second Term)} + \mbox{\rm (Third Term)} + 7\beta \cdot H,
\end{align}
where
\begin{align*}
\mbox{\rm (First Term)}
& \eqdef \mbox{$r_{\jdistr} \cdot \sqrt{8\beta \cdot \jdistri[1](r_{\jdistr}) \cdot \big(1 - \jdistri[1](r_{\jdistr})\big)}$}. \\
\mbox{\rm (Second Term)}
& \eqdef \mbox{$\int_{r_{\jdistr}}^{\max\{r_{\jdistr},~\mathrm{B}\}} \sqrt{8\beta \cdot \jdistri[2](x) \cdot \big(1 - \jdistri[2](x)\big)} \cdot \d x$}. \\
\mbox{\rm (Third Term)}
& \eqdef \mbox{$\int_{\max\{r_{\jdistr},~\mathrm{B}\}}^{H} \sqrt{8\beta \cdot \jdistri[2](x) \cdot \big(1 - \jdistri[2](x)\big)} \cdot \d x$}.
\end{align*}
We measure these terms in the next two lemmas.

\begin{lemma}
\label{lem:1_H:1}
$\mbox{\rm (First Term)} + \mbox{\rm (Second Term)} \leq \sqrt{8\beta \cdot H} \cdot \ar(r_{\jdistr},~\jdistr) = \sqrt{8\beta \cdot H} \cdot \ar(\jdistr)$.
\end{lemma}

\begin{proof}
Recall \Cref{fact:1_H:2} that $r_{\jdistr} \leq \jdistri[1]^{-1}(\frac{H - 1}{H})$, which implies $\jdistri[1](r_{\jdistr}) \leq \frac{H - 1}{H}$ and thus $\frac{\jdistri[1](r_{\jdistr})}{1 - \jdistri[1](r_{\jdistr})} \leq H - 1 \leq H$. Consequently,
\begin{align*}
\mbox{\rm (First Term)}
& = \mbox{$\sqrt{8\beta \cdot \frac{\jdistri[1](r_{\jdistr})}{1 - \jdistri[1](r_{\jdistr})}} \cdot r_{\jdistr} \cdot \big(1 - \jdistri[1](r_{\jdistr})\big)$} \\
& \leq \mbox{$\sqrt{8\beta \cdot H} \cdot r_{\jdistr} \cdot \big(1 - \jdistri[1](r_{\jdistr})\big)$}.
\end{align*}
Similarly, $\sqrt{\jdistri[2](\val) \cdot \big(1 - \jdistri[2](\val)\big)} \leq \sqrt{H} \cdot \big(1 - \jdistri[2](\val)\big)$ whenever $\val \leq \mathrm{B} = {\jdistri[2]}^{-1}(\frac{H - 1}{H})$. Hence,\footnote{Particularly, even if $r_{\jdistr} \geq \mathrm{B}$, we still have $\mbox{\rm (Second Term)} = 0 \leq \sqrt{8\beta \cdot H} \cdot \int_{r_{\jdistr}}^{H} \big(1 - \jdistri[2](x)\big) \cdot \d x$.}
\begin{align*}
\mbox{\rm (Second Term)}
& = \mbox{$\int_{r_{\jdistr}}^{\max\{r_{\jdistr},~\mathrm{B}\}} \sqrt{8\beta \cdot \jdistri[2](x) \cdot \big(1 - \jdistri[2](x)\big)} \cdot \d x$} \\
& \leq \mbox{$\sqrt{8\beta \cdot H} \cdot \int_{r_{\jdistr}}^{\max\{r_{\jdistr},~\mathrm{B}\}} \big(1 - \jdistri[2](x)\big) \cdot \d x$} \\
& \leq \mbox{$\sqrt{8\beta \cdot H} \cdot \int_{r_{\jdistr}}^{H} \big(1 - \jdistri[2](x)\big) \cdot \d x$}.
\end{align*}
Combining the above two inequalities together completes the proof of Lemma~\ref{lem:1_H:1}.
\end{proof}

\begin{lemma}
\label{lem:1_H:2}
$\mbox{\rm (Third Term)} \leq \sqrt{8\beta \cdot H}$.
\end{lemma}

\begin{proof}
Clearly, the second-highest CDF $\jdistri[2](\val) \leq 1$ for any value $\val \in \RRP$. For any value $\val \geq \mathrm{B} = {\jdistri[2]}^{-1}(\frac{H - 1}{H}) \in [1, H]$, we have $1 - \jdistri[2](\val) \leq \frac{1}{H}$. Accordingly,
\begin{align*}
\mbox{\rm (Third Term)}
& = \mbox{$\int_{\max\{r_{\jdistr},~\mathrm{B}\}}^{H} \sqrt{8\beta \cdot \jdistri[2](x) \cdot \big(1 - \jdistri[2](x)\big)}$} \cdot \d x \\
& \leq \mbox{$\int_{\max\{r_{\jdistr},~\mathrm{B}\}}^{H} \sqrt{8\beta / H}$} \cdot \d x
\leq \mbox{$H \cdot \sqrt{8\beta / H}$}
= \mbox{$\sqrt{8\beta \cdot H}$}.
\end{align*}
This completes the proof of Lemma~\ref{lem:1_H:2}.
\end{proof}

\noindent
Applying Lemmas~\ref{lem:1_H:1} and \ref{lem:1_H:2} to inequality~\eqref{eq:1_H:0}, we conclude that $\ar(\sdistr) \geq (1 - \eps) \cdot \ar(\jdistr)$:
\begin{align*}
\ar(\jdistr) - \ar(\sdistr)
& \leq \mbox{$\sqrt{8\beta \cdot H} \cdot \ar(\jdistr) + \sqrt{8\beta \cdot H} + 7\beta \cdot H$} \\
& \leq \mbox{$(2 \cdot \sqrt{8\beta \cdot H} + 7\beta \cdot H) \cdot \ar(\jdistr)$}
&& \hspace{-1.15cm} \mbox{\tt (\Cref{fact:1_H:1}: $\ar(\jdistr) \geq 1$)} \\
& \leq \mbox{$(\frac{\sqrt{6}}{3} \cdot \eps + \frac{7}{48} \cdot \eps^2) \cdot \ar(\jdistr)$}
&& \hspace{-1.15cm} \mbox{\tt (Part~2 of \Cref{fact:parameter}: $\beta \geq \frac{\eps^2 \cdot H^{-1}}{48}$)} \\
& \leq \mbox{$\eps \cdot \ar(\jdistr)$}.
&& \hspace{-1.15cm} \mbox{\tt (as $\frac{\sqrt{6}}{3} + \frac{7}{48} \approx 0.9623 < 1$)}
\end{align*}
This completes the proof in the $[1, H]$-bounded setting.

\subsection{Continuous Regular Setting}
\label{subsec:revenue:regular}

Throughout this subsection, we assume that each buyer $j \in [n]$ {\em independently} draws his value (for the item) from a {\em continuous regular} distribution $\distri$. Different from the former two settings, a regular distribution may have an {\em unbounded support}, which incurs extra technical challenges in proving the desired sample complexity of \Cref{alg:main}.

To address this issue, we carefully truncate the given instance $\jdistr$, such that (1)~the resulting instance $\tdistr$ is still close to $\jdistr$, under the measurement of the optimal {\arm} revenue; (2)~$\tdistr$ has a small enough support supremum, which allows us to bound the revenue gap between it and its shaded counterpart $\stdistr$ ($\grave{a}$ la the proofs in the former two settings). Indeed, (3)~$\stdistr$ is dominated by the shaded instance $\sdistr$ (derived directly from $\jdistr$), thus $\ar(\stdistr) \leq \ar(\jdistr)$. Combining everything together completes the proof in this setting.

\vspace{.1in}
\noindent
{\bf Auxiliary Lemmas.}
To elaborate the truncation scheme, let us introduce several useful facts. Below, Lemma~\ref{lem:first_second} might be known in the literature, yet we include a short proof for completeness. Notably, it only requires the distributions $\jdistr = \{\distri\}_{j \in [n]}$ to be independent.

\begin{lemma}[Order Statistics]
\label{lem:first_second}
For any product instance $\jdistr = \{\distri\}_{j \in [n]}$, the highest CDF $\jdistri[1]$ and the second-highest CDF $\jdistri[2]$ satisfy that $1 - \jdistri[2](\val) \leq \big(1 - \jdistri[1](\val)\big)^2$ for any value $\val \in \RRP$.
\end{lemma}

\begin{proof}
After elementary algebra (see \cite[Section~4]{JLTX20}), one can easily check that the highest CDF $\jdistri[1](\val) = \prod_{j \in [n]} \distri(\val)$ and the second-highest CDF
\begin{align*}
\jdistri[2](\val)
& = \mbox{$\sum_{i \in [n]} \Pr\big[\samplei[i] \geq \val \wedge (\samplei < \val, \forall j \neq i)\big]$} \\
& \phantom{=} \qquad\qquad + \mbox{$\Pr[\samplei < \val,~\forall j \in [n]]$}
&& \mbox{\tt (draw $\{\samplei\}_{j = 1}^{n}$ from $\{\distri\}_{i \in [n]}$)} \\
& = \mbox{$\jdistri[1](\val) \cdot \big[1 + \sum_{j \in [n]} \big(1 / \distri(\val) - 1\big)\big]$} \\
& \geq \mbox{$\jdistri[1](\val) \cdot \big[1 + \sum_{j \in [n]} \ln\big(1 / \distri(\val)\big)\big]$}
&& \mbox{\tt (as $z \geq \ln(1 + z)$ when $z \in \RRP$)} \\
& = \jdistri[1](\val) \cdot \big(1 - \ln\jdistri[1](\val)\big)
&& \mbox{\tt (as $\jdistri[1](\val) = \prod_{j \in [n]} \distri(\val)$)} \\
& \geq \jdistri[1](\val) \cdot \big(2 - \jdistri[1](\val)\big).
&& \mbox{\tt (as $\ln(1 - z) \leq -z$ when $z \in [0, 1]$)}
\end{align*}
We thus conclude the proof of Lemma~\ref{lem:first_second} by rearranging the above inequality.
\end{proof}

We safely scale the original instance $\jdistr = \{\distri\}_{j \in [n]}$ so that $\max_{\val \in \RRP} \big\{\val \cdot \big(1 - \jdistri[1](\val)\big)\big\} = 1$. Together with Lemma~\ref{lem:first_second}, this normalization leads to the following observations.

\begin{fact}
\label{fact:regular:ar}
$\ar(\jdistr) = \max_{r \in \RRP} \big\{r \cdot \big(1 - \jdistri[1](r)\big) + \int_{r}^{\infty} \big(1 - \jdistri[2](x)\big) \cdot \d x\big\} \geq 1$.
\end{fact}

\begin{fact}
\label{fact:regular:1}
The highest CDF $\jdistri[1]$ is stochastically dominated by the equal-revenue CDF $\Phi_1$, namely $\jdistri[1](\val) \geq \Phi_1(\val) \eqdef (1 - \frac{1}{\val})_+$ for any value $\val \in \RRP$.
\end{fact}

\begin{fact}
\label{fact:regular:2}
The second-highest CDF $\jdistri[2]$ is stochastically dominated by the CDF $\Phi_2(\val) \eqdef (1 - \frac{1}{\val^2})_+$, namely $\jdistri[2](\val) \geq \Phi_2(\val)$ for any value $\val \in \RRP$.
\end{fact}

\noindent
{\bf Truncation Scheme.}
Based on the original instance $\jdistr = \jdistri[1] \uplus \jdistri[2]$, we construct the truncated instance $\tdistr = \tdistri[1] \uplus \tdistri[2]$ as follows: for both $i \in \{1,~2\}$ and any value $\val \in \RRP$,
\begin{align}
\label{eq:construction:regular}\tag{\rm Truncation}
& \tdistri(\val) \eqdef
\begin{cases}
\jdistri(\val) & \mbox{when $\jdistri(\val) \leq 1 - (\eps / 4)^i$} \\
1 & \mbox{when $\jdistri(\val) > 1 - (\eps / 4)^i$}
\end{cases}.
\end{align}
We immediately get two useful facts about the truncated instance $\tdistr = \tdistri[1] \uplus \tdistri[2]$.

\begin{fact}
\label{fact:regular:3}
For $i \in \{1,~2\}$, the truncated $i$-th highest CDF $\tdistri$ is dominated by the original $i$-th highest CDF $\jdistri$. Thus, the shaded counterpart $\stdistri = \shade(\tdistri)$ is dominated by $\sdistri = \shade(\jdistri)$.
\end{fact}

\begin{proof}
The first dominance $\tdistri \bedominated \jdistri$ is obvious (by construction). The second dominance $\stdistri \bedominated \sdistri$ also holds, because $\shade(\cdot)$ is a non-decreasing function (see \Cref{fact:shade_function}).
\end{proof}

\begin{fact}
\label{fact:regular:4}
The truncated instance $\tdistr = \tdistri[1] \uplus \tdistri[2]$ has a support supremum of $s_u \leq 4 / \eps$.
\end{fact}

\begin{proof}
As we certified in Lemma~\ref{lem:first_second}, for any value $\val \in \RRP$, the highest and second-highest CDF's satisfy that $1 - \jdistri[2](\val) \leq \big(1 - \jdistri[1](\val)\big)^2$. From this one can derive that
\begin{align*}
\mbox{${\jdistri[2]}^{-1}(1 - \eps^2 / 16) \leq {\jdistri[1]}^{-1}(1 - \eps / 4)$}.
\end{align*}
For each $i$-th highest CDF $\jdistri$, we indeed truncate the particular $(\frac{\eps}{4})^i$-fraction of quantiles that correspond to the largest possible values. In view of the above inequality, the truncated second-highest CDF $\tdistri[2]$ must have a smaller support supremum than the truncated highest CDF $\tdistri[1]$. Due to \Cref{fact:regular:1}, we further have $\Phi_1(s_u) \leq \jdistri[1](s_u) = \tdistri[1](s_u) \leq 1 - \eps / 4$. That is, $1 - 1 / s_u \leq 1 - \eps / 4$ and thus $s_u \leq 4 / \eps$. This completes the proof of \Cref{fact:regular:4}.
\end{proof}

\noindent
{\bf Revenue Loss.}
Below, Lemma~\ref{lem:regular:1} shows that \eqref{eq:construction:regular} only incurs a small revenue loss.

\begin{lemma}[Revenue Loss]
\label{lem:regular:1}
The truncated instance $\tdistr = \tdistri[1] \uplus \tdistri[2]$ satisfies that
\[
\mbox{$\ar(\tdistr) \geq (1 - \frac{3}{4} \cdot \eps) \cdot \ar(\jdistr)$}.
\]
\end{lemma}

\begin{proof}
We adopt a hybrid argument. For brevity, let $\ar(r,~\jdistri[1] \uplus \tdistri[2])$ be the resulting {\arm} revenue (under any reserve $r \in \RRP$) when only the second-highest CDF is truncated, and let $\overline{r}$ be the optimal reserve for the {\em hybrid} instance $\jdistri[1] \uplus \tdistri[2]$. The lemma comes from these two inequalities:
\begin{align}
\label{eq:lem:regular:1:2}
\ar(\tdistr)
& \geq (1 - \eps / 4) \cdot \ar(\jdistri[1] \uplus \tdistri[2]) \\
\label{eq:lem:regular:1:3}
& \geq (1 - \eps / 4) \cdot (1 - \eps / 2) \cdot \ar(\jdistr).
\end{align}
In the remainder of the proof, we verify these two inequalities one by one.

\vspace{.1in}
\noindent
{\bf Inequality~\eqref{eq:lem:regular:1:2}.}
Under replacing the original highest CDF $\jdistri[1]$ with $\tdistri[1]$, we claim that
\begin{align}
\label{eq:lem:regular:1:4}\tag{$\star$}
\exists (r \leq \overline{r}): \qquad r \cdot \big(1 - \tdistri[1](r)\big)
\geq (1 - \eps / 4) \cdot \overline{r} \cdot \big(1 - \jdistri[1](\overline{r})\big).
\end{align}
The new reserve $r \in [0,~\overline{r}]$ may not be optimal for the truncated instance $\tdistr = \tdistri[1] \uplus \tdistri[2]$. Based on the revenue formula and assuming inequality~\eqref{eq:lem:regular:1:4}, we can infer inequality~\eqref{eq:lem:regular:1:2}:
\begin{align*}
\ar(\tdistr)
& \geq \ar(r,~\tdistr)
= \mbox{$r \cdot \big(1 - \tdistri[1](r)\big) + \int_{r}^{\infty} \big(1 - \tdistri[2](x)\big) \cdot \d x$} \\
& \geq \mbox{$r \cdot \big(1 - \tdistri[1](r)\big) + (1 - \eps / 4) \cdot \int_{\overline{r}}^{\infty} \big(1 - \tdistri[2](x)\big) \cdot \d x$}
\hspace{1cm} \mbox{\tt (as $r \leq \overline{r}$)} \\
& \geq (1 - \eps / 4) \cdot \ar(\overline{r},~\jdistri[1] \uplus \tdistri[2])
\hspace{3.38cm} \mbox{\tt (inequality~\eqref{eq:lem:regular:1:4})} \\
& = (1 - \eps / 4) \cdot \ar(\jdistri[1] \uplus \tdistri[2]).
\hspace{1.7cm} \mbox{\tt ($\overline{r}$ is optimal for $\jdistri[1] \uplus \tdistri[2]$)}
\end{align*}
It remains to verify inequality~\eqref{eq:lem:regular:1:4}. If $\jdistri[1](\overline{r}) < 1 - \eps / 4$, by construction we have $\tdistri[1](\val) = \jdistri[1](\val)$ for any value $\val \leq \overline{r}$. Clearly, inequality~\eqref{eq:lem:regular:1:4} holds by employing the same reserve $r \gets \overline{r}$.

From now on, we safely assume $\jdistri[1](\overline{r}) = \prod_{j \in [n]} \distri(\overline{r}) \geq 1 - \eps / 4$. Inequality~\eqref{eq:lem:regular:1:4} is enabled by the next \Cref{fact:triangular}, which can be summarized from \cite[Section~2]{AHNPY19}.

\begin{fact}
\label{fact:triangular}
For any continuous regular distribution $\distri$ and any value $\overline{r} \in \RRP$, define the parameter $a_j \eqdef \overline{r} \cdot \big(1 / \distri(\overline{r}) - 1\big)$. Then, $\distri(\val) \leq \frac{\val}{\val + a_j}$ for any value $\val \in [0,~\overline{r}]$, with the equality holds when $\val = \overline{r}$.
\end{fact}

\noindent
Consider another auxiliary highest CDF $\mathcal{G}_1^*(\val) \eqdef \prod_{j \in [n]} \frac{\val}{\val + a_j}$. In view of \Cref{fact:triangular}, it suffices to show the following instead of inequality~\eqref{eq:lem:regular:1:4}:
\begin{align}
\label{eq:lem:regular:1:5}\tag{$\diamond$}
\exists (r \leq \overline{r}): \qquad r \cdot \big(1 - \mathcal{G}_1^*(r)\big) \geq (1 - \eps / 4) \cdot \overline{r} \cdot \big(1 - \jdistri[1](\overline{r})\big),
\end{align}
We choose $r \gets \mathcal{G}_1^{*-1}(1 - \eps / 4)$. Since $\mathcal{G}_1^*(\overline{r}) = \jdistri[1](\overline{r}) \geq 1 - \eps / 4$ (by \Cref{fact:triangular} and our assumption) and $\mathcal{G}_1^*$ is an increasing function, we do have $r \leq \overline{r}$. Let us bound the new reserve $r$ from below:
\begin{align*}
\mbox{$1 - \eps / 4
= \mathcal{G}_1^*(r)
= \prod_{j \in [n]} \frac{r}{r + a_j}
\leq \frac{r}{r + \sum_{j \in [n]} a_j}$}
\qquad\Rightarrow\qquad
\mbox{$r \geq (4 / \eps - 1) \cdot \sum_{j \in [n]} a_j$}.
\end{align*}
Given this, we can accomplish inequality~\eqref{eq:lem:regular:1:5} as follows:
\begin{align*}
\lhs \mbox{ of \eqref{eq:lem:regular:1:5}}
& = \mbox{$r \cdot (\eps / 4)$}
\geq \mbox{$(1 - \eps / 4) \cdot \sum_{j \in [n]} a_j$} \\
& = \mbox{$(1 - \eps / 4) \cdot \overline{r} \cdot \sum_{j \in [n]} \big(1 / \distri(\overline{r}) - 1\big)$}
\hspace{1.12cm} \mbox{\tt (by definition of $a_j$)} \\
& \geq \mbox{$(1 - \eps / 4) \cdot \overline{r} \cdot \sum_{j \in [n]} \big(1 - \distri(\overline{r})\big)$}
\hspace{1.5cm} \mbox{\tt (as CDF $\distri(\overline{r}) \in [0, 1]$)} \\
& \geq \mbox{$(1 - \eps / 4) \cdot \overline{r} \cdot \big(1 - \prod_{j \in [n]} \distri(\overline{r})\big)$}
= \rhs \mbox{ of \eqref{eq:lem:regular:1:5}},
\end{align*}
where the last inequality is because $\sum z_i \geq 1 - \prod (1 - z_i)$ when $z_i$'s are between $[0, 1]$.

\vspace{.1in}
\noindent
{\bf Inequality~\eqref{eq:lem:regular:1:3}.}
Since the reserve $r_{\jdistr}$ is optimal for the original instance $\jdistr = \jdistri[1] \uplus \jdistri[2]$ but may not for the hybrid instance $\jdistri[1] \uplus \tdistri[2]$, we deduce from the revenue formula that
\begin{align*}
\ar(\jdistr) - \ar(\jdistri[1] \uplus \tdistri[2])
& \leq \ar(r_{\jdistr},~\jdistri[1] \uplus \jdistri[2]) - \ar(r_{\jdistr},~\jdistri[1] \uplus \tdistri[2]) \\
& = \mbox{$\int_{r_{\jdistr}}^{\infty} \big(\tdistri[2](x) - \jdistri[2](x)\big) \cdot \d x$}.
\end{align*}
By construction, $0 \leq \tdistri[2](\val) - \jdistri[2](\val) \leq (\frac{\eps}{4})^2$ for any value $\val \in \RRP$. Also, it follows from \Cref{fact:regular:2} that $\jdistri[2](\val) + \frac{1}{\val^2} \geq 1 \geq \tdistri[2](\val)$. Apply both facts to the $\rhs$ of the above inequality:
\begin{align*}
\ar(\jdistr) - \ar(\jdistri[1] \uplus \tdistri[2])
& \leq \mbox{$\int_{0}^{\infty} \big(\tdistri[2](x) - \jdistri[2](x)\big) \cdot \d x$}
&& \mbox{\tt (lengthen the interval)} \\
& \leq \mbox{$\int_{0}^{\infty} \min\big\{(\frac{\eps}{4})^2,~\frac{1}{x^2}\big\}$} \cdot \d x \\
& = \eps / 2 \leq (\eps / 2) \cdot \ar(\jdistr),
&& \mbox{\tt (\Cref{fact:regular:ar}: $\ar(\jdistr) \geq 1$)}
\end{align*}
which gives inequality~\eqref{eq:lem:regular:1:3} after rearranging. This completes the proof of Lemma~\ref{lem:regular:1}.
\end{proof}

We now prove that, when the sample complexity $m \geq 11520\eps^{-3} \cdot (\ln\eps^{-1} + \ln\delta^{-1} + 4)$, the optimal {\arm} revenue from the shaded truncated instance $\stdistr = \stdistri[1] \uplus \stdistri[2]$ is indeed close enough to that from the truncated instance $\tdistr = \tdistri[1] \uplus \tdistri[2]$.

\begin{lemma}
\label{lem:regular:2}
The following holds for the shaded truncated instance $\stdistr = \stdistri[1] \uplus \stdistri[2]$:
\begin{align*}
\ar(\stdistr) \geq \ar(\tdistr) - \eps / 4.
\end{align*}
\end{lemma}

\begin{proof}
Denote by $r^*$ the optimal reserve for the truncated instance $\tdistr = \tdistri[1] \uplus \tdistri[2]$. Clearly, $r^*$ is at most the support supremum of $s_u \leq 4 / \eps$ (see \Cref{fact:regular:4}), and may not be optimal for the shaded truncated instance $\stdistr = \stdistri[1] \uplus \stdistri[2]$. As illustrated in the former two settings, the function $\shade(x) \leq x + \sqrt{8\beta \cdot x \cdot (1 - x)} + 7\beta$ for any $x \in [0, 1]$. Given these,\footnote{Note that the interval of integration can be safely truncated to the support supremum of $s_u \leq 4 / \eps$.}
\begin{align}
\notag
\ar(\tdistr) - \ar(\stdistr)
& \leq \ar(r^*,~\tdistr) - \ar(r^*,~\stdistr)
\quad \mbox{\tt ($r^*$ may not be optimal to $\stdistr$)} \\
\notag
& = \mbox{$r^* \cdot \big(\shade(\tdistri[1](r^*)) - \tdistri[1](r^*)\big) + \int_{r^*}^{4 / \eps} \big(\shade(\tdistri[2](x)) - \tdistri[2](x)\big) \cdot \d x$} \\
\label{eq:regular:0}
& \leq \mbox{\rm (First Term)} + \mbox{\rm (Second Term)} + 28\beta \cdot \eps^{-1},
\end{align}
where
\begin{align*}
\mbox{\rm (First Term)}
& \eqdef \mbox{$r^* \cdot \sqrt{8\beta \cdot \jdistri[1](r^*) \cdot \big(1 - \jdistri[1](r^*)\big)}$}. \\
\mbox{\rm (Second Term)}
& \eqdef \mbox{$\int_{r^*}^{4 / \eps} \sqrt{8\beta \cdot \tdistri[2](x) \cdot \big(1 - \tdistri[2](x)\big)} \cdot \d x$}.
\end{align*}
In the reminder of the proof, we quantify these two terms one by one.

\vspace{.1in}
\noindent
{\bf First Term.}
We infer from \Cref{fact:regular:1,fact:regular:3} that the truncated highest CDF $\tdistri[1](\val) \geq 1 - \frac{1}{\val}$ for any value $\val \in \RRP$. Additionally, of course $\tdistri[1](\val) \leq 1$. We thus have
\begin{align*}
\mbox{\rm (First Term)}
& \leq \mbox{$r^* \cdot \sqrt{8\beta \cdot 1 \cdot \big[1 - (1 - 1 / r^*)\big]}$} \\
& = \mbox{$\sqrt{8\beta \cdot r^*}$}
\leq \mbox{$\sqrt{32\beta \cdot \eps^{-1}}$}.
&& \mbox{\tt (as $r^* \leq s_u \leq 4 / \eps$)}
\end{align*}

\vspace{.1in}
\noindent
{\bf Second Term.}
Based on \Cref{fact:regular:2,fact:regular:3}, for any value $\val \in \RRP$, the truncated second-highest CDF $\tdistri[2](\val) \geq (1 - \frac{1}{\val^2})_+$. Also, of course $\tdistri[2](\val) \leq 1$. For these reasons,
\begin{align*}
\mbox{\rm (Second Term)}
& \leq \mbox{$\int_{0}^{4 / \eps} \sqrt{8\beta \cdot \big(1 - \tdistri[2](x)\big)} \cdot \d x$}
&& \mbox{\tt (as $\tdistri[2](x) \leq 1$)} \\
& \leq \mbox{$\int_{0}^{1} \sqrt{8\beta} \cdot \d x + \int_{1}^{4 / \eps} \sqrt{8\beta \cdot \frac{1}{x^2}}$} \cdot \d x
&& \mbox{\tt (as $\tdistri[2](x) \geq (1 - \frac{1}{x^2})_+$)} \\
& = \mbox{$\sqrt{8\beta} + \sqrt{8\beta} \cdot \ln(4 / \eps)$}
= \mbox{$\sqrt{8\beta} \cdot \ln(4e / \eps)$}
\end{align*}
Plug the above two inequalities into inequality~\eqref{eq:regular:0}:
\begin{align*}
\ar(\tdistr) - \ar(\stdistr)
& \leq \mbox{$\sqrt{\frac{32\beta}{\eps}} + \sqrt{8\beta} \cdot \ln(\frac{4e}{\eps}) + \frac{28\beta}{\eps}$} \\
& \leq \mbox{$\frac{\eps}{\sqrt{90}} + \frac{\eps^{3 / 2} \cdot \ln(4e / \eps)}{\sqrt{360}} + \frac{7\eps^2}{720}$}
&& \mbox{\tt (Part~3 of \Cref{fact:parameter}: $\beta \leq \frac{\eps^3}{2880}$)} \\
& \leq \mbox{$\frac{\eps}{\sqrt{90}} + \frac{\ln(4e) \cdot \eps}{\sqrt{360}} + \frac{7\eps^2}{720}$}
&& \mbox{\tt ($\sqrt{\eps} \cdot \ln(\frac{4e}{\eps}) \leq \ln(4e)$ for $0 < \eps < 1$)} \\
& \leq \mbox{$\eps / 4$}
&& \mbox{\tt ($\frac{1}{\sqrt{90}} + \frac{\ln(4e)}{\sqrt{360}} + \frac{7}{720} \approx 0.2409 < \frac{1}{4}$)}
\end{align*}
This completes the proof of Lemma~\ref{lem:regular:2}.
\end{proof}

The next Corollary~\ref{crl:regular} accomplishes the proof in the continuous regular setting.

\begin{corollary}
\label{crl:regular}
When the sample complexity $m \geq 11520\eps^{-3} \cdot (\ln\eps^{-1} + \ln\delta^{-1} + 4)$:
\begin{align*}
\ar(\sdistr)
& \geq \ar(\stdistr)
&& \mbox{\tt (\Cref{fact:regular:3}: dominance $\sdistri \dominate \stdistri$)} \\
& \geq \mbox{$\ar(\tdistr) - \eps / 4$}
&& \mbox{\tt (Lemma~\ref{lem:regular:2}: $\ar(\stdistr) \geq \ar(\tdistr) - \eps / 4$)} \\
& \geq \mbox{$\ar(\tdistr) - (\eps / 4) \cdot \ar(\jdistr)$}
&& \mbox{\tt (\Cref{fact:regular:ar}: $\ar(\jdistr) \geq 1$)} \\
& \geq (1 - \eps) \cdot \ar(\jdistr).
&& \mbox{\tt (Lemma~\ref{lem:regular:1}: $\ar(\tdistr) \geq (1 - \frac{3}{4} \cdot \eps) \cdot \ar(\jdistr)$)}
\end{align*}
\end{corollary}

\subsection{\texorpdfstring{$\mhr$}{} Setting}
\label{subsec:revenue:mhr}

In this subsection, we also assume that the original distributions $\jdistr = \{\distri\}_{j \in [n]}$ are independent, and scale the instance such that $\max_{\val \in \RRP} \big\{\val \cdot \big(1 - \jdistri[1](\val)\big)\big\} = 1$. Therefore, Lemma~\ref{lem:first_second} and \Cref{fact:regular:ar,fact:regular:1,fact:regular:2} still holds. Nevertheless, the lower-bound formulas in \Cref{fact:regular:1,fact:regular:2} (for the highest and second-highest CDF's) actually have too {\em heavy tails}. Namely, {\em sharper} formulas are required to prove the desired revenue gap between the original instance $\jdistr$ and its shaded counterpart $\sdistr$, given the more demanding sample complexity of $m = \cO\big(\eps^{-2} \cdot (\ln\eps^{-1} + \ln\delta^{-1})\big)$.

Based on the particular structures of the $\mhr$ distributions, we will first obtain workable lower-bound formulas, and then quantify the revenue loss between $\ar(\sdistr)$ and $\ar(\jdistr)$. To this end, we safely assume $m \geq 5610\eps^{-2} \cdot (\ln\eps^{-1} + \ln\delta^{-1} + 5)$.

\vspace{.1in}
\noindent
{\bf Lower-Bound CDF Formulas.}
Below, Lemma~\ref{lem:mhr:3} shows that the highest and second-highest CDF's of any $\mhr$ instance decay exponentially fast.

\begin{lemma}
\label{lem:mhr:3}
The following holds for any continuous or discrete $\mhr$ instance $\jdistr = \{\distri\}_{j \in [n]}$:
\begin{enumerate}[font = {\em \bfseries}]
\item The highest CDF $\jdistri[1](\val) \geq 1 - \frac{3}{2} \cdot e^{-\val / 6}$ for any value $\val \geq e$.
\item The second-highest CDF $\jdistri[2](\val) \geq 1 - \frac{9}{4} \cdot e^{-\val / 3}$ for any value $\val \geq e$.
\item The shaded instance $\sdistr = \sdistri[1] \uplus \sdistri[2]$ has a support supremum of $s_u \leq 12\ln(\frac{21}{\eps})$.
\end{enumerate}
\end{lemma}

\begin{proof}
To see {\bfseries Item~1}, we fix a parameter $u > 1$ (to be determined) and present a reduction (from the original $\mhr$ distributions $\jdistr = \{\distri\}_{j \in [n]}$ to certain {\em continuous exponential distributions}) such that, for any value $\val \geq u$, the highest CDF decreases point-wise.

\begin{figure}[htbp]
\centering
\subfloat[Discrete $\mhr$ distribution
\label{fig:lem:mhr:3:1}]{
\begin{tikzpicture}[thick, smooth, scale = 2.75]
\draw[->] (0, 0) -- (2.1, 0);
\draw[->] (0, -1.5) -- (0, 0.15);
\node[above] at (0, 0.15) {\small $\mathrm{y}$};
\node[right] at (2.1, 0) {\small $\val$};
\node[left] at (0, 0) {\small $0$};
\node[anchor = north east, color = blue] at (1.8856, -1.275) {\small $\mathrm{G}_j(\val) = \ln\big(1 - \distri(\val)\big)$};
\draw[color = blue] (0, 0) -- (0.4714, 0) (0.4714, -0.0833) -- (0.9428, -0.0833) (0.9428, -0.3333) -- (1.4142, -0.3333) (1.4142, -0.75) -- (1.8856, -0.75) (1.8856, -1.3333) -- (2, -1.3333);
\draw[densely dotted, color = blue] (0.4714, 0) -- (0.4714, -0.0833) (0.9428, -0.0833) -- (0.9428, -0.3333) (1.4142, -0.3333) -- (1.4142, -0.75) (1.8856, -0.75) -- (1.8856, -1.3333);
\draw[color = darkgray!80] (0.4714, 0) -- (0.9428, -0.0833) -- (1.4142, -0.3333) -- (1.8856, -0.75) -- (2.1, -1.0154);
\draw[color = red] (0, 0) -- (2, -0.4715);
\node[above] at (1.4142, 0) {\small $u$};
\draw[dashed] (1.4142, -0.3333) -- (1.4142, 0);
\draw[dashed] (0.9428, -0.3333) -- (0, -0.3333);
\draw[very thick] (1.4142, 0) -- (1.4142, -1pt);
\draw[very thick] (0, -0.3333) -- (1pt, -0.3333);
\draw[color = black, fill = green] (0.4714, 0) circle(0.5pt);
\draw[color = black, fill = green] (0.9428, -0.0833) circle(0.5pt);
\draw[color = black, fill = green] (1.4142, -0.3333) circle(0.5pt);
\draw[color = black, fill = green] (1.8856, -0.75) circle(0.5pt);
\draw[color = black, fill = white] (0.4714, -0.0833) circle(0.5pt);
\draw[color = black, fill = white] (0.9428, -0.3333) circle(0.5pt);
\draw[color = black, fill = white] (1.4142, -0.75) circle(0.5pt);
\draw[color = black, fill = white] (1.8856, -1.3333) circle(0.5pt);
\draw[color = black, fill = green] (0, 0) circle(0.5pt);
\end{tikzpicture}}
\hspace{1cm}
\subfloat[Continuous $\mhr$ distribution
\label{fig:lem:mhr:3:2}]{
\begin{tikzpicture}[thick, smooth, scale = 2.75]
\draw[->] (0, 0) -- (2.1, 0);
\draw[->] (0, -1.5) -- (0, 0.15);
\node[above] at (0, 0.15) {\small $\mathrm{y}$};
\node[right] at (2.1, 0) {\small $\val$};
\node[left] at (0, 0) {\small $0$};
\node[anchor = north east, color = blue] at (1.8856, -1.275) {\small $\mathrm{G}_j(\val) = \ln\big(1 - \distri(\val)\big)$};
\draw[color = blue, domain = 0: 1.8856] plot (\x, {-3 * \x^2 / 8});
\draw[color = red] (0, 0) -- (2, -1.0607);
\node[above] at (1.4142, 0) {\small $u$};
\draw[dashed] (1.4142, -0.75) -- (1.4142, 0);
\draw[dashed] (1.4142, -0.75) -- (0, -0.75);
\draw[very thick] (1.4142, 0) -- (1.4142, -1pt);
\draw[very thick] (0, -0.75) -- (1pt, -0.75);
\draw[color = black, fill = green] (0, 0) circle(0.5pt);
\draw[color = black, fill = green] (1.4142, -0.75) circle(0.5pt);
\end{tikzpicture}}
\caption{Demonstration for the reduction in the proof of \Cref{lem:mhr:3}.}
\label{fig:lem:mhr:3}
\end{figure}

We first handle the {\em discrete} $\mhr$ instances. As \Cref{fig:lem:mhr:3:1} illustrates and by definition (see \Cref{subsec:prelim:distr}), such an instance $\jdistr = \{\distri\}_{j \in [n]}$ has a discrete support of $\{k \cdot \Delta: k \in \NNP\}$, where the step-size $\Delta > 0$ is fixed. We must have $\Delta \leq 1$, because the instance is scaled so that $\max_{\val \in \RRP} \big\{\val \cdot \big(1 - \jdistri[1](\val)\big)\big\} = 1$ and $\Delta$ is exactly the support infimum (i.e., $\jdistri[1](\Delta) = 0$).

For any $j \in [n]$, let us consider the {\em step function} $\mathrm{G}_j(\val) \eqdef \ln\big(1 - \distri(\val)\big)$ (marked in blue in \Cref{fig:lem:mhr:3:1}) and the {\em piece-wise linear function} $\mathrm{L}_j$ (marked in gray) induced by the origin $(0,~0)$ and the {\em ``\raisebox{-1ex}{\LARGE $\urcorner$}''-type points} $\big(k \cdot \Delta,~\mathrm{G}_j(k \cdot \Delta)\big)$'s (marked in green). Apparently, $\mathrm{G}_j(\val) \leq \mathrm{L}_j(\val)$ for any value $\val \in \RRP$.

The $\mhr$ condition holds iff $\mathrm{L}_j$ is a {\em concave function} (see \Cref{subsec:prelim:distr}). Choose $u \gets k \cdot \Delta$ (for some $k \in \NNP$ to be determined) and let $a_j \eqdef -\frac{1}{u} \cdot \mathrm{G}_j(u) > 0$, we infer from \Cref{fig:lem:mhr:3:1}:
\begin{align}
\label{eq:lem:mhr:3:1}
-a_j \cdot \val \geq \mathrm{L}_j(\val) \geq \mathrm{G}_j(\val) = \ln\big(1 - \distri(\val)\big)
\qquad\Rightarrow\qquad
\distri(\val) \geq 1 - e^{-a_j \cdot \val},
\end{align}
for any value $\val \geq u$, with all the equalities holding when $\val = u$. Given these, we also have
\begin{align}
\label{eq:lem:mhr:3:2}
\mbox{$-a_j \cdot u = \ln\big(1 - \distri(u)\big) \leq \ln\big(1 - \prod_{j \in [n]} \distri(u)\big) = \ln\big(1 - \jdistri[1](u)\big)$},
\end{align}
for each $j \in [n]$. Put everything together: for any value $\val \geq u$,
\begin{align*}
\ln\jdistri[1](\val)
& = \mbox{$\sum_{j \in [n]} \ln\distri(\val)$}
&& \mbox{\tt (as $\jdistri[1](\val) = \prod_{j \in [n]} \distri(\val)$)} \\
& \overset{\eqref{eq:lem:mhr:3:1}}{\geq} \mbox{$\sum_{j \in [n]} \ln(1 - e^{-a_j \cdot \val})$} \\
& = \mbox{$-\sum_{j \in [n]} \sum_{p = 1}^{\infty} \frac{1}{p} \cdot e^{-p \cdot a_j \cdot u} \cdot e^{-p \cdot a_j \cdot u \cdot (\val / u - 1)}$}
&& \mbox{\tt (Taylor series)} \\
& \overset{\eqref{eq:lem:mhr:3:2}}{\geq} \mbox{$-e^{\ln(1 - \jdistri[1](u)) \cdot (\val / u - 1)} \cdot \sum_{j \in [n]} \sum_{p = 1}^{\infty} \frac{1}{p} \cdot e^{-p \cdot a_j \cdot u}$}
&& \mbox{\tt ($p \geq 1$ and $\val / u - 1 \geq 0$)} \\
& = \mbox{$\big(1 - \jdistri[1](u)\big)^{\val / u - 1} \cdot \ln\prod_{j \in [n]} (1 - e^{-a_j \cdot u})$}
&& \mbox{\tt (Taylor series)} \\
& \overset{\eqref{eq:lem:mhr:3:1}}{=} \big(1 - \jdistri[1](u)\big)^{\val / u - 1} \cdot \ln\jdistri[1](u),
&& \mbox{\tt (equality condition)}
\end{align*}
from which we deduce that $\jdistri[1](\val) \geq \big(\jdistri[1](u)\big)^{(1 - \jdistri[1](u))^{\val / u - 1}}$ for any value $\val \geq u = k \cdot \Delta$. It can be seen that this lower-bound formula is an increasing function in the term $\jdistri[1](u) \in [0, 1]$.

We would like to choose $k \gets \lfloor e / \Delta\rfloor$. Because the step-size $\Delta \leq 1$, we do have $k \in \NNP$ and $u = k \cdot \Delta \in [e - 1,~e]$. Then, it follows from \Cref{fact:regular:1} that $\jdistri[1](u) \geq \jdistri[1](e - 1) \geq 1 - \frac{1}{e - 1}$. Replace the term $\jdistri[1](u)$ in the above lower-bound formula with this bound:
\begin{align*}
\jdistri[1](\val)
& \geq \mbox{$\big(\frac{e - 2}{e - 1}\big)^{(e - 1)^{1 - \val / u}}$}
= \mbox{$e^{-\ln(\frac{e - 1}{e - 2}) \cdot (e - 1)^{1 - \val / u}}$} \\
& \geq \mbox{$1 - \ln(\frac{e - 1}{e - 2}) \cdot (e - 1)^{1 - \val / u}$}
&& \mbox{\tt (as $e^{-z} \geq 1 - z$)} \\
& \geq \mbox{$1 - \ln(\frac{e - 1}{e - 2}) \cdot (e - 1)^{1 - \val / e}$}
&& \mbox{\tt (as $u \leq e$)} \\
& = \mbox{$1 - (e - 1) \cdot \ln(\frac{e - 1}{e - 2}) \cdot e^{-\frac{\ln(e - 1)}{e} \cdot \val}$} \\
& \geq \mbox{$1 - \frac{3}{2} \cdot e^{-\val / 6}$},
&& \mbox{\tt (elementary algebra)}
\end{align*}
for any value $\val \in [u,~\infty)$. Clearly, this inequality holds in the shorter range of $\val \in [e,~\infty)$.

When $\jdistr = \{\distri\}_{j \in [n]}$ is a {\em continuous} $\mhr$ instance, by definition (see \Cref{subsec:prelim:distr}) each function $\mathrm{G}_j(\val) = \ln\big(1 - \distri(\val)\big)$ itself is a concave function (as \Cref{fig:lem:mhr:3:2} illustrates). That is, we can simply choose $u \gets e$ and apply the same arguments as the above. Actually, we can get a better lower-bound formula that $\jdistri[1](\val) \geq 1 - \frac{5}{4} \cdot e^{-\val / e}$ for any value $\val \geq e$.

Clearly, {\bfseries Item~2} is an implication of {\bfseries Item~1} and Lemma~\ref{lem:first_second}. Now, we turn to attesting {\bfseries Item~3}. By definition, the function $\shade(x) = \min \big\{1,~~x + \sqrt{8\beta \cdot x \cdot (1 - x)} + 7\beta\big\} = 1$ when $x \geq 1 - 7\beta$. Hence, the shaded instance $\sdistr = \sdistri[1] \uplus \sdistri[2]$ has a support supremum of
\begin{align*}
s_u
& \leq \mbox{$\max \{{\jdistri[1]}^{-1}(1 - 7\beta),~~{\jdistri[2]}^{-1}(1 - 7\beta)\}$}
&& \mbox{\tt (dominance $\jdistri[1] \dominate \jdistri[2]$)} \\
& = \mbox{${\jdistri[1]}^{-1}(1 - 7\beta)$}
\leq \mbox{$6\ln(\frac{3}{14} \cdot \beta^{-1})$}
&& \mbox{\tt (Part~1 of Lemma~\ref{lem:mhr:3})} \\
& \leq \mbox{$6\ln(\frac{2805}{7\eps^2})$}
\leq \mbox{$12\ln(21 / \eps)$}.
&& \mbox{\tt (Part~4 of \Cref{fact:parameter}: $\beta \leq \frac{\eps^2}{1870}$)}
\end{align*}
This completes the proof of Lemma~\ref{lem:mhr:3}.
\end{proof}

\noindent
{\bf Revenue Loss.}
Conceivably, the original instance $\jdistr = \jdistri[1] \uplus \jdistri[2]$ should have a  small optimal reserve $r_{\jdistr}$, since $\jdistri[1]$ and $\jdistri[2]$ both have {\em light tails}. This proposition is formalized as the next Lemma~\ref{lem:mhr:4}, which will be useful in our later proof.

\begin{lemma}
\label{lem:mhr:4}
For the original $\mhr$ instance $\jdistr = \jdistri[1] \uplus \jdistri[2]$, there is an optimal {\arm} auction having a reserve of $r_{\jdistr} \leq \mathcal{C}^*$, where the constant $\mathcal{C}^* \approx 20.5782$ is the larger one between the two roots of the transcendental equation $\frac{3}{2} \cdot z \cdot e^{-z / 6} = 1$.
\end{lemma}

\begin{proof}
The proof here is similar in spirit to that of \Cref{fact:1_H:2}. When there are multiple alternative optimal reserves $r_{\jdistr}$'s, we would select the {\em smallest} one. To see the lemma, we need the math fact $\frac{3}{2} \cdot z \cdot e^{-z / 6} < 1$ when $z > \mathcal{C}^* \approx 20.5782$. Then, it follows from Part~1 of Lemma~\ref{lem:mhr:3} that
\begin{align}
\label{eq:lem:mhr:4:0}
\mbox{$r \cdot \big(1 - \jdistri[1](r)\big) \leq \frac{3}{2} \cdot r \cdot e^{-r / 6} < 1$},
\end{align}
for any reserve $r > \mathcal{C}^*$. Particularly, $\lim_{r \to \infty} r \cdot \big(1 - \jdistri[1](r)\big) = 0$. By contrast, we have scaled the instance such that $\max_{\val \in \RRP} \big\{\val \cdot \big(1 - \jdistri[1](\val)\big)\big\} = 1$, which means $\overline{r} \cdot \big(1 - \jdistri[1](\overline{r})\big) = 1$ for some other reserve $\overline{r} \in [0,~\mathcal{C}^*]$. Recall the {\arm} revenue formula:
\begin{align*}
\ar(\overline{r},~\jdistr) - \ar(r,~\jdistr)
& = \mbox{$1 - r \cdot \big(1 - \jdistri[1](r)\big) + \int_{\overline{r}}^{r} \big(1 - \jdistri[2](x)\big) \cdot \d x$} \\
& \geq \mbox{$1 - r \cdot \big(1 - \jdistri[1](r)\big)
\overset{\eqref{eq:lem:mhr:4:0}}{\geq} 0$}.
\hspace{1cm} \mbox{\tt (as $\overline{r} \leq \mathcal{C}^* < r$)}
\end{align*}
That is, under our tie-breaking rule, any reserve $r > \mathcal{C}^*$ cannot be revenue-optimal. Apparently, this observation indicates Lemma~\ref{lem:mhr:4}.
\end{proof}

Finally, Lemma~\ref{lem:mhr:2} establishes the desired revenue gap between the original instance $\jdistr = \jdistri[1] \uplus \jdistri[2]$ and its shaded counterpart $\sdistr = \sdistri[1] \uplus \sdistri[2]$, thus settling the $\mhr$ case.

\begin{lemma}
\label{lem:mhr:2}
When the sample complexity $m \geq 5610\eps^{-2} \cdot (\ln\eps^{-1} + \ln\delta^{-1} + 5)$:
\begin{align*}
\ar(\sdistr) \geq (1 - \eps) \cdot \ar(\jdistr).
\end{align*}
\end{lemma}

\begin{proof}
Recall that the function $\shade(x) \leq x + \sqrt{8\beta \cdot x \cdot (1 - x)} + 7\beta$ when $x \in [0, 1]$. Based on the support supremum $s_u \leq 12\ln(\frac{21}{\eps})$ established in Part~3 of Lemma~\ref{lem:mhr:3} and the {\arm} revenue formula, we deduce that\footnote{Note that the interval of integration can be safely truncated to the support supremum of $s_u \leq 12\ln(\frac{21}{\eps})$.}
\begin{align}
\notag
\ar(\jdistr) - \ar(\sdistr)
& \leq \ar(r_{\jdistr},~\jdistr) - \ar(r_{\jdistr},~\sdistr)
\hspace{1cm} \mbox{\tt ($r_{\jdistr}$ may not be optimal to $\sdistr$)} \\
\notag
& = \mbox{$r_{\jdistr} \cdot \big(\sdistri[1](r_{\jdistr}) - \jdistri[1](r_{\jdistr})\big) + \int_{r_{\jdistr}}^{12\ln(21 / \eps)} \big(\sdistri[2](x) - \jdistri[2](x)\big) \cdot \d x$} \\
\notag
& \phantom{=} \mbox{$\phantom{r_{\jdistr} \cdot \big(\sdistri[1](r_{\jdistr}) - \jdistri[1](r_{\jdistr})\big)} + \int_{12\ln(21 / \eps)}^{\infty} \big(1 - \jdistri[2](x)\big) \cdot \d x$} \\
\label{eq:mhr:0}
& \leq \mbox{\rm (First Term)} + \mbox{\rm (Second Term)} + \mbox{\rm (Third Term)} + 84\beta \cdot \ln(21 / \eps),
\end{align}
where
\begin{align*}
\mbox{\rm (First Term)}
& \eqdef \mbox{$r_{\jdistr} \cdot \sqrt{8\beta \cdot \jdistri[1](r_{\jdistr}) \cdot \big(1 - \jdistri[1](r_{\jdistr})\big)}$}. \\
\mbox{\rm (Second Term)}
& \eqdef \mbox{$\int_{r_{\jdistr}}^{12\ln(21 / \eps)} \sqrt{8\beta \cdot \jdistri[2](x) \cdot \big(1 - \jdistri[2](x)\big)} \cdot \d x$}. \\
\mbox{\rm (Third Term)}
& \eqdef \mbox{$\int_{12\ln(21 / \eps)}^{\infty} \big(1 - \jdistri[2](x)\big) \cdot \d x$}.
\end{align*}
In the reminder of the proof, we quantify these three terms one by one.

\vspace{.1in}
\noindent
\textcolor{blue!75}{\rm [First Term].}
Recall \Cref{fact:regular:1} that the highest CDF $\jdistri[1](\val) \geq 1 - \frac{1}{\val}$ for any value $\val \in \RRP$. Further, of course $\jdistri[1](\val) \leq 1$. Given these and because $r_{\jdistr} \leq \mathcal{C}^* \approx 20.5782$ (see Lemma~\ref{lem:mhr:4}),
\begin{align*}
\mbox{\rm (First Term)}
\leq \mbox{$r_{\jdistr} \cdot \sqrt{8\beta \cdot 1 \cdot \big[1 - (1 - 1 / r_{\jdistr})\big]}$}
= \mbox{$\sqrt{8\beta \cdot r_{\jdistr}}$}
\leq \mbox{$\sqrt{165\beta}$}.
\end{align*}

\noindent
\textcolor{blue!75}{\rm [Second Term].}
Clearly, $\jdistri[2](\val) \in [0, 1]$ for all value $\val \in \RRP$. Additionally, we infer from Part~2 of Lemma~\ref{lem:mhr:3} that $\sqrt{1 - \jdistri[2](\val)} \leq \frac{3}{2} \cdot e^{-\val / 6}$ when $\val \geq e$. For these reasons,
\begin{align*}
\mbox{\rm (Second Term)}
& \leq \mbox{$\int_{0}^{\infty} \sqrt{8\beta \cdot \jdistri[2](x) \cdot \big(1 - \jdistri[2](x)\big)} \cdot \d x$} \\
& \leq \mbox{$\int_{0}^{e} \sqrt{8\beta} \cdot \d x + \int_{e}^{\infty} \sqrt{8\beta} \cdot \frac{3}{2} \cdot e^{-x / 6} \cdot \d x$} \\
& = \mbox{$\sqrt{8\beta} \cdot (e + 9e^{-e / 6})$}
\leq \mbox{$\sqrt{570\beta}$}
&&\hspace{-13pt} \mbox{\tt (elementary algebra)}
\end{align*}

\noindent
\textcolor{blue!75}{\rm [Third Term].}
Also, we deduce from Part~2 of \Cref{lem:mhr:3} that
\begin{align*}
\mbox{\rm (Third Term)}
= \mbox{$\int_{12\ln(21 / \eps)}^{\infty} \big(1 - \jdistri[2](x)\big) \cdot \d x$}
\leq \mbox{$\int_{12\ln(21 / \eps)}^{\infty} \frac{9}{4} \cdot e^{-x / 3} \cdot \d x$}
= \mbox{$\frac{\varepsilon^{4}}{28812}$}
\leq \mbox{$\frac{\varepsilon}{28812}$}.
\end{align*}

\noindent
Plug the above three inequalities into inequality~\eqref{eq:mhr:0}:
\begin{align*}
\ar(\jdistr) - \ar(\sdistr)
& \leq \mbox{$\sqrt{165\beta} + \sqrt{570\beta} + \frac{\varepsilon}{28812} + 84\beta \cdot \ln(21 / \eps)$} \\
& \leq \mbox{$\sqrt{\frac{3}{34}} \eps + \sqrt{\frac{57}{187}} \eps + \frac{\varepsilon}{28812} + \frac{42\eps^2 \cdot \ln(21 / \eps)}{935}$}
&& \hspace{-.32cm} \mbox{\tt (\Cref{fact:parameter}, Part~4: $\beta \leq \frac{\eps^2}{1870}$)} \\
& \leq \mbox{$\sqrt{\frac{3}{34}} \eps + \sqrt{\frac{57}{187}} \eps + \frac{\varepsilon}{28812} + \frac{42\ln21}{935} \eps$}
&& \hspace{-.32cm} \mbox{\tt ($\eps \cdot \ln(\frac{21}{\eps}) \leq \ln21$ for $0 < \eps < 1$)} \\
& \leq \mbox{$\eps$}
\leq \mbox{$\eps \cdot \ar(\jdistr)$}
&&\hspace{-13pt} \mbox{\tt (elementary algebra)}
\end{align*}
where the last inequality is due to \Cref{fact:regular:ar}. This completes the proof of Lemma~\ref{lem:mhr:2}.
\end{proof}

\subsection{Continuous \texorpdfstring{$\lambda$-Regular}{} Setting}

In the literature, there is another distribution family that receives much attention \cite{CN91,CR14,CR17,ABB22} -- the continuous $\lambda$-regular distributions. When the built-in parameter $\lambda$ ranges from $0$ to $1$, this family smoothly expands from the $\mhr$ family to the regular family.

$\grave{A}$ la the $\mhr$ case, the sample complexity upper bound is still $\cO\big(\eps^{-2} \cdot (\ln\eps^{-1} + \ln\delta^{-1})\big)$, despite that the $\cO(\cdot)$ notation now hides some absolute constant $\mathcal{C}_{\lambda}$ depending on $\lambda \in (0, 1)$. Since the proof of this bound is very similar to the $\mhr$ case, we just show in \Cref{app:extension} a counterpart extreme value theorem (cf.\ Lemma~\ref{lem:mhr:3}), but omit the other parts about the revenue smoothness analysis.

It is noteworthy that the $\tO\big(\eps^{-2}\big)$ upper bound may not be optimal. Namely, in both of the continuous $\lambda$-regular setting and the continuous $\mhr$ setting, the best known lower bounds are $\cOmega(\eps^{-3 / 2})$ \cite{HMR18}. It would be interesting to pin down the exact sample complexity in both settings, for which the tools developed here and by \cite{CD15,HMR18,GHZ19} might be useful.

\section{Conclusion and Further Discussion}
\label{sec:conclusion}

In this work, we proved the nearly tight sample complexity of the {\arm} auction,
for each of the $[0, 1]^{n}$-bounded, $[1, H]^{n}$-bounded, regular and $\mhr$ distribution families. In the literature on ``mechanism design via sampling'', a notion complementary to sample complexity is {\em regret minimization} (e.g., see \cite{BHW02,BKRW03,BH05} and the follow-up papers). Regarding the {\arm} auction, this means the seller must select a careful reserve price $r_{t} \in \RRP$ in each round $t$ over a time horizon $\mathrm{T} \in \NNP$, in order to maximize the {\em cumulative revenue}, i.e., minimize the cumulative revenue loss against a certain benchmark.

Indeed, if the seller can access the highest and second-highest bids in all of the past $(t - 1)$ rounds, our results imply the nearly optimal regret bounds. Consider the $[0, 1]$-additive setting for example. Because $\cO(\eps^{-2} \cdot \ln\eps^{-1})$ samples suffice to reduce the revenue loss to $\eps \in (0, 1)$, the regret in each round $t \in [\mathrm{T}]$ is at most $\cO(\sqrt{(\ln t) / t})$. As a result, the cumulative regret is at most $\sum_{t = 1}^{\mathrm{T}} \cO(\sqrt{(\ln t) / t}) = \cO(\sqrt{\mathrm{T} \cdot \ln\mathrm{T}})$. Similarly, the $\cOmega(\eps^{-2})$ lower bound on the sample complexity implies an $\cOmega(\sqrt{\mathrm{T}})$ lower bound on the regret.

\cite{CGM15} considered the same problem under {\em weaker data access}, where the seller can only observe the allocations and the payments in the past $(t - 1)$ rounds. This models some particular markets, where the seller is not the auctioneer and can acquire a least amount of information. Assuming the bids are i.i.d.\ and supported on $[0, 1]$, Cesa-Bianchi et al.\ proved a matching regret of $\tTheta(\sqrt{\mathrm{T}})$.\footnote{More precisely, their upper bound is $\cO(\sqrt{\mathrm{T} \cdot \ln\ln\ln\mathrm{T}} \cdot \ln\ln\mathrm{T})$ and their lower bound is also $\cOmega(\sqrt{\mathrm{T}})$.} But, what if the seller still has the weak data access yet the distributions are distinct and even correlated? The regret of the {\arm} auction and other mechanisms in this model is an interesting problem.

Additionally, another natural and meaningful adjusted model is to assume that the bidders would {\em strategically} report their samples, or further, that the bidders themselves are {\em learners} as well. At the time of our paper, this research direction is very nascent yet has already received much attention. For an overview of this, the reader can turn to \cite{BDHN19,HLW18,HT19,FL20} and the references therein. 

\vspace{.1in}
\noindent
{\bf Acknowledgement.}
We would like to thank Zhiyi Huang, Xi Chen, Rocco Servedio, and anonymous reviewers for many helpful discussions and comments.

Y.J.\ is supported by NSF grants IIS-1838154, CCF-1563155, CCF-1703925, CCF-1814873, CCF-2106429, and CCF-2107187.
P.L.\ is supported by Science and Technology Innovation 2030 – ``New Generation of Artificial Intelligence'' Major Project No.(2018AAA0100903), NSFC grant 61922052 and 61932002, Innovation Program of Shanghai Municipal Education Commission, Program for Innovative Research Team of Shanghai University of Finance and Economics, and Fundamental Research Funds for Central Universities.

\bibliographystyle{alpha}
\bibliography{main}

\newcommand{\etalchar}[1]{$^{#1}$}
\begin{thebibliography}{BGMM18}

\bibitem[ABB22]{ABB22}
Amine Allouah, Achraf Bahamou, and Omar Besbes.
\newblock Pricing with samples.
\newblock {\em Oper. Res.}, 70(2):1088--1104, 2022.

\bibitem[AHN{\etalchar{+}}19]{AHNPY19}
Saeed Alaei, Jason~D. Hartline, Rad Niazadeh, Emmanouil Pountourakis, and Yang
  Yuan.
\newblock Optimal auctions vs. anonymous pricing.
\newblock {\em Games Econ. Behav.}, 118:494--510, 2019.

\bibitem[AM06]{LMP06}
Lawrence~M Ausubel and Paul Milgrom.
\newblock The lovely but lonely vickrey auction.
\newblock {\em Combinatorial auctions}, 17:22--26, 2006.

\bibitem[BBHM08]{BBHM08}
Maria{-}Florina Balcan, Avrim Blum, Jason~D. Hartline, and Yishay Mansour.
\newblock Reducing mechanism design to algorithm design via machine learning.
\newblock {\em J. Comput. Syst. Sci.}, 74(8):1245--1270, 2008.

\bibitem[BDHN19]{BDHN19}
S{\'{e}}bastien Bubeck, Nikhil~R. Devanur, Zhiyi Huang, and Rad Niazadeh.
\newblock Multi-scale online learning: Theory and applications to online
  auctions and pricing.
\newblock {\em J. Mach. Learn. Res.}, 20:62:1--62:37, 2019.

\bibitem[Ber24]{B24}
Sergei Bernstein.
\newblock On a modification of chebyshev's inequality and of the error formula
  of laplace.
\newblock {\em Ann. Sci. Inst. Sav. Ukraine, Sect. Math}, 1(4):38--49, 1924.

\bibitem[BGLT19]{BGLT19}
Xiaohui Bei, Nick Gravin, Pinyan Lu, and Zhihao~Gavin Tang.
\newblock Correlation-robust analysis of single item auction.
\newblock In {\em Proceedings of the Thirtieth Annual {ACM-SIAM} Symposium on
  Discrete Algorithms, {SODA} 2019, San Diego, California, USA, January 6-9,
  2019}, pages 193--208, 2019.

\bibitem[BGMM18]{BGMM18}
Moshe Babaioff, Yannai~A. Gonczarowski, Yishay Mansour, and Shay Moran.
\newblock Are two (samples) really better than one?
\newblock In {\em Proceedings of the 2018 {ACM} Conference on Economics and
  Computation, Ithaca, NY, USA, June 18-22, 2018}, page 175, 2018.

\bibitem[BH05]{BH05}
Avrim Blum and Jason~D. Hartline.
\newblock Near-optimal online auctions.
\newblock In {\em Proceedings of the Sixteenth Annual {ACM-SIAM} Symposium on
  Discrete Algorithms, {SODA} 2005, Vancouver, British Columbia, Canada,
  January 23-25, 2005}, pages 1156--1163, 2005.

\bibitem[BHW02]{BHW02}
Ziv Bar{-}Yossef, Kirsten Hildrum, and Felix Wu.
\newblock Incentive-compatible online auctions for digital goods.
\newblock In {\em Proceedings of the Thirteenth Annual {ACM-SIAM} Symposium on
  Discrete Algorithms, January 6-8, 2002, San Francisco, CA, {USA.}}, pages
  964--970, 2002.

\bibitem[BKRW03]{BKRW03}
Avrim Blum, Vijay Kumar, Atri Rudra, and Felix Wu.
\newblock Online learning in online auctions.
\newblock In {\em Proceedings of the Fourteenth Annual {ACM-SIAM} Symposium on
  Discrete Algorithms, January 12-14, 2003, Baltimore, Maryland, {USA.}}, pages
  202--204, 2003.

\bibitem[BMP63]{BMP63}
Richard~E. Barlow, Albert~W. Marshall, and Frank Proschan.
\newblock Properties of probability distributions with monotone hazard rate.
\newblock {\em The Annals of Mathematical Statistics}, 34(2):375--389, 1963.

\bibitem[BSV16]{BSV16}
Maria{-}Florina Balcan, Tuomas Sandholm, and Ellen Vitercik.
\newblock Sample complexity of automated mechanism design.
\newblock In {\em Advances in Neural Information Processing Systems 29: Annual
  Conference on Neural Information Processing Systems 2016, December 5-10,
  2016, Barcelona, Spain}, pages 2083--2091, 2016.

\bibitem[BSV18]{BSV18}
Maria{-}Florina Balcan, Tuomas Sandholm, and Ellen Vitercik.
\newblock A general theory of sample complexity for multi-item profit
  maximization.
\newblock In {\em Proceedings of the 2018 {ACM} Conference on Economics and
  Computation, Ithaca, NY, USA, June 18-22, 2018}, pages 173--174, 2018.

\bibitem[CD15]{CD15}
Yang Cai and Constantinos Daskalakis.
\newblock Extreme value theorems for optimal multidimensional pricing.
\newblock {\em Games and Economic Behavior}, 92:266--305, 2015.

\bibitem[CD17]{CD17}
Yang Cai and Constantinos Daskalakis.
\newblock Learning multi-item auctions with (or without) samples.
\newblock In {\em 58th {IEEE} Annual Symposium on Foundations of Computer
  Science, {FOCS} 2017, Berkeley, CA, USA, October 15-17, 2017}, pages
  516--527, 2017.

\bibitem[CDO{\etalchar{+}}22]{CDOPSY22}
Xi~Chen, Ilias Diakonikolas, Anthi Orfanou, Dimitris Paparas, Xiaorui Sun, and
  Mihalis Yannakakis.
\newblock On the complexity of optimal lottery pricing and randomized
  mechanisms for a unit-demand buyer.
\newblock {\em {SIAM} J. Comput.}, 51(3):492--548, 2022.

\bibitem[CDP{\etalchar{+}}18]{CDPSY18}
Xi~Chen, Ilias Diakonikolas, Dimitris Paparas, Xiaorui Sun, and Mihalis
  Yannakakis.
\newblock The complexity of optimal multidimensional pricing for a unit-demand
  buyer.
\newblock {\em Games and Economic Behavior}, 110:139--164, 2018.

\bibitem[CGM15]{CGM15}
Nicol{\`{o}} Cesa{-}Bianchi, Claudio Gentile, and Yishay Mansour.
\newblock Regret minimization for reserve prices in second-price auctions.
\newblock {\em {IEEE} Trans. Information Theory}, 61(1):549--564, 2015.

\bibitem[CHLW11]{CHLW11}
Xue Chen, Guangda Hu, Pinyan Lu, and Lei Wang.
\newblock On the approximation ratio of k-lookahead auction.
\newblock In {\em Internet and Network Economics - 7th International Workshop,
  {WINE} 2011, Singapore, December 11-14, 2011. Proceedings}, pages 61--71,
  2011.

\bibitem[CHMS10]{CHMS10}
Shuchi Chawla, Jason~D. Hartline, David~L. Malec, and Balasubramanian Sivan.
\newblock Multi-parameter mechanism design and sequential posted pricing.
\newblock In {\em Proceedings of the 42nd {ACM} Symposium on Theory of
  Computing, {STOC} 2010, Cambridge, Massachusetts, USA, 5-8 June 2010}, pages
  311--320, 2010.

\bibitem[CMPY18]{CMPY18}
Xi~Chen, George Matikas, Dimitris Paparas, and Mihalis Yannakakis.
\newblock On the complexity of simple and optimal deterministic mechanisms for
  an additive buyer.
\newblock In {\em Proceedings of the Twenty-Ninth Annual {ACM-SIAM} Symposium
  on Discrete Algorithms, {SODA} 2018, New Orleans, LA, USA, January 7-10,
  2018}, pages 2036--2049, 2018.

\bibitem[CN91]{CN91}
Andrew Caplin and Barry Nalebuff.
\newblock Aggregation and social choice: a mean voter theorem.
\newblock {\em Econometrica: Journal of the Econometric Society}, pages 1--23,
  1991.

\bibitem[CR14]{CR14}
Richard Cole and Tim Roughgarden.
\newblock The sample complexity of revenue maximization.
\newblock In {\em Symposium on Theory of Computing, {STOC} 2014, New York, NY,
  USA, May 31 - June 03, 2014}, pages 243--252, 2014.

\bibitem[CR17]{CR17}
Richard Cole and Shravas Rao.
\newblock Applications of {\(\alpha\)}-strongly regular distributions to
  bayesian auctions.
\newblock {\em {ACM} Trans. Economics and Comput.}, 5(4):18:1--18:29, 2017.

\bibitem[DDT14]{DDT14}
Constantinos Daskalakis, Alan Deckelbaum, and Christos Tzamos.
\newblock The complexity of optimal mechanism design.
\newblock In {\em Proceedings of the Twenty-Fifth Annual {ACM-SIAM} Symposium
  on Discrete Algorithms, {SODA} 2014, Portland, Oregon, USA, January 5-7,
  2014}, pages 1302--1318, 2014.

\bibitem[DFK15]{DFK15}
Shahar Dobzinski, Hu~Fu, and Robert Kleinberg.
\newblock Approximately optimal auctions for correlated bidders.
\newblock {\em Games and Economic Behavior}, 92:349--369, 2015.

\bibitem[DHP16]{DHP16}
Nikhil~R. Devanur, Zhiyi Huang, and Christos{-}Alexandros Psomas.
\newblock The sample complexity of auctions with side information.
\newblock In {\em Proceedings of the 48th Annual {ACM} {SIGACT} Symposium on
  Theory of Computing, {STOC} 2016, Cambridge, MA, USA, June 18-21, 2016},
  pages 426--439, 2016.

\bibitem[DRY15]{DRY15}
Peerapong Dhangwatnotai, Tim Roughgarden, and Qiqi Yan.
\newblock Revenue maximization with a single sample.
\newblock {\em Games Econ. Behav.}, 91:318--333, 2015.

\bibitem[DS22]{DS22}
Constantinos Daskalakis and Vasilis Syrgkanis.
\newblock Learning in auctions: Regret is hard, envy is easy.
\newblock {\em Games Econ. Behav.}, 134:308--343, 2022.

\bibitem[FILS15]{FILS15}
Hu~Fu, Nicole Immorlica, Brendan Lucier, and Philipp Strack.
\newblock Randomization beats second price as a prior-independent auction.
\newblock In {\em Proceedings of the Sixteenth {ACM} Conference on Economics
  and Computation, {EC} '15, Portland, OR, USA, June 15-19, 2015}, page 323,
  2015.

\bibitem[FL20]{FL20}
Hu~Fu and Tao Lin.
\newblock Learning utilities and equilibria in non-truthful auctions.
\newblock In Hugo Larochelle, Marc'Aurelio Ranzato, Raia Hadsell,
  Maria{-}Florina Balcan, and Hsuan{-}Tien Lin, editors, {\em Advances in
  Neural Information Processing Systems 33: Annual Conference on Neural
  Information Processing Systems 2020, NeurIPS 2020, December 6-12, 2020,
  virtual}, 2020.

\bibitem[GHTZ21]{GHTZ21}
Chenghao Guo, Zhiyi Huang, Zhihao~Gavin Tang, and Xinzhi Zhang.
\newblock Generalizing complex hypotheses on product distributions: Auctions,
  prophet inequalities, and pandora's problem.
\newblock In Mikhail Belkin and Samory Kpotufe, editors, {\em Conference on
  Learning Theory, {COLT} 2021, 15-19 August 2021, Boulder, Colorado, {USA}},
  volume 134 of {\em Proceedings of Machine Learning Research}, pages
  2248--2288. {PMLR}, 2021.

\bibitem[GHZ19]{GHZ19}
Chenghao Guo, Zhiyi Huang, and Xinzhi Zhang.
\newblock Settling the sample complexity of single-parameter revenue
  maximization.
\newblock In {\em Proceedings of the 51st Annual {ACM} {SIGACT} Symposium on
  Theory of Computing, {STOC} 2019, Phoenix, AZ, USA, June 23-26, 2019.}, pages
  662--673, 2019.

\bibitem[GN17]{GN17}
Yannai~A. Gonczarowski and Noam Nisan.
\newblock Efficient empirical revenue maximization in single-parameter auction
  environments.
\newblock In {\em Proceedings of the 49th Annual {ACM} {SIGACT} Symposium on
  Theory of Computing, {STOC} 2017, Montreal, QC, Canada, June 19-23, 2017},
  pages 856--868, 2017.

\bibitem[GW21]{GW21}
Yannai~A. Gonczarowski and S.~Matthew Weinberg.
\newblock The sample complexity of up-to-{\(\epsilon\)} multi-dimensional
  revenue maximization.
\newblock {\em J. {ACM}}, 68(3):15:1--15:28, 2021.

\bibitem[Har13]{H13}
Jason~D Hartline.
\newblock Mechanism design and approximation.
\newblock {\em Book draft. October}, 122, 2013.

\bibitem[HMR18]{HMR18}
Zhiyi Huang, Yishay Mansour, and Tim Roughgarden.
\newblock Making the most of your samples.
\newblock {\em {SIAM} J. Comput.}, 47(3):651--674, 2018.

\bibitem[HR09]{HR09}
Jason~D. Hartline and Tim Roughgarden.
\newblock Simple versus optimal mechanisms.
\newblock In {\em Proceedings 10th {ACM} Conference on Electronic Commerce
  (EC-2009), Stanford, California, USA, July 6--10, 2009}, pages 225--234,
  2009.

\bibitem[HT19]{HT19}
Jason~D. Hartline and Samuel Taggart.
\newblock Sample complexity for non-truthful mechanisms.
\newblock In {\em Proceedings of the 2019 {ACM} Conference on Economics and
  Computation, {EC} 2019, Phoenix, AZ, USA, June 24-28, 2019.}, pages 399--416,
  2019.

\bibitem[JJLZ22]{JJLZ22}
Yaonan Jin, Shunhua Jiang, Pinyan Lu, and Hengjie Zhang.
\newblock Tight revenue gaps among multiunit mechanisms.
\newblock {\em SIAM Journal on Computing}, 51(5):1535--1579, 2022.

\bibitem[JLQ19a]{JLQ19}
Yaonan Jin, Weian Li, and Qi~Qi.
\newblock On the approximability of simple mechanisms for {MHR} distributions.
\newblock In Ioannis Caragiannis, Vahab~S. Mirrokni, and Evdokia Nikolova,
  editors, {\em Web and Internet Economics - 15th International Conference,
  {WINE} 2019, New York, NY, USA, December 10-12, 2019, Proceedings}, volume
  11920 of {\em Lecture Notes in Computer Science}, pages 228--240. Springer,
  2019.

\bibitem[JLQ{\etalchar{+}}19b]{JLQTX19}
Yaonan Jin, Pinyan Lu, Qi~Qi, Zhihao~Gavin Tang, and Tao Xiao.
\newblock Tight approximation ratio of anonymous pricing.
\newblock In {\em Proceedings of the 51st Annual {ACM} {SIGACT} Symposium on
  Theory of Computing, {STOC} 2019, Phoenix, AZ, USA, June 23-26, 2019.}, pages
  674--685, 2019.

\bibitem[JLTX20]{JLTX20}
Yaonan Jin, Pinyan Lu, Zhihao~Gavin Tang, and Tao Xiao.
\newblock Tight revenue gaps among simple mechanisms.
\newblock {\em {SIAM} J. Comput.}, 49(5):927--958, 2020.

\bibitem[LHW18]{HLW18}
Jinyan Liu, Zhiyi Huang, and Xiangning Wang.
\newblock Learning optimal reserve price against non-myopic bidders.
\newblock In {\em Advances in Neural Information Processing Systems 31: Annual
  Conference on Neural Information Processing Systems 2018, NeurIPS 2018, 3-8
  December 2018, Montr{\'{e}}al, Canada.}, pages 2042--2052, 2018.

\bibitem[MM16]{MM16}
Mehryar Mohri and Andres~Mu{\~{n}}oz Medina.
\newblock Learning algorithms for second-price auctions with reserve.
\newblock {\em Journal of Machine Learning Research}, 17:74:1--74:25, 2016.

\bibitem[MR15]{MR15}
Jamie Morgenstern and Tim Roughgarden.
\newblock On the pseudo-dimension of nearly optimal auctions.
\newblock In {\em Advances in Neural Information Processing Systems 28: Annual
  Conference on Neural Information Processing Systems 2015, December 7-12,
  2015, Montreal, Quebec, Canada}, pages 136--144, 2015.

\bibitem[MR16]{MR16}
Jamie Morgenstern and Tim Roughgarden.
\newblock Learning simple auctions.
\newblock In {\em Proceedings of the 29th Conference on Learning Theory, {COLT}
  2016, New York, USA, June 23-26, 2016}, pages 1298--1318, 2016.

\bibitem[Mye81]{M81}
Roger~B. Myerson.
\newblock Optimal auction design.
\newblock {\em Math. Oper. Res.}, 6(1):58--73, 1981.

\bibitem[PP15]{PP15}
Christos~H. Papadimitriou and George Pierrakos.
\newblock Optimal deterministic auctions with correlated priors.
\newblock {\em Games and Economic Behavior}, 92:430--454, 2015.

\bibitem[Ron01]{R01}
Amir Ronen.
\newblock On approximating optimal auctions.
\newblock In {\em Proceedings 3rd {ACM} Conference on Electronic Commerce
  (EC-2001), Tampa, Florida, USA, October 14-17, 2001}, pages 11--17, 2001.

\bibitem[RS16]{RS16}
Tim Roughgarden and Okke Schrijvers.
\newblock Ironing in the dark.
\newblock In {\em Proceedings of the 2016 {ACM} Conference on Economics and
  Computation, {EC} '16, Maastricht, The Netherlands, July 24-28, 2016}, pages
  1--18, 2016.

\bibitem[Syr17]{S17}
Vasilis Syrgkanis.
\newblock A sample complexity measure with applications to learning optimal
  auctions.
\newblock In {\em Advances in Neural Information Processing Systems 30: Annual
  Conference on Neural Information Processing Systems 2017, 4-9 December 2017,
  Long Beach, CA, {USA}}, pages 5352--5359, 2017.

\end{thebibliography}

\appendix
\renewcommand{\appendixname}{Appendix~\Alph{section}}

\section{Missing Proof in \texorpdfstring{\Cref{sec:prelim,sec:sample}}{}}
\label{app:sample}

For any technical result and its proof to be presented in this appendix, the reader can find the counterparts from \cite[Appendix~B]{GHZ19}.

\subsection{Proof of \texorpdfstring{\Cref{fact:shade_function}}{}}
\label{subapp:fact:shade_function}

\noindent\fbox{
\begin{minipage}{0.977\textwidth}
{\bf \Cref{fact:shade_function}.}
{\em The following two functions are both non-decreasing functions on interval $x \in [0, 1]$.
\begin{itemize}
\item $\shade(x) = \min \big\{1,~~x + \sqrt{8\beta \cdot x \cdot (1 - x)} + 7\beta\big\}$
\item $\eshade(x) = \min \big\{1,~~x + \sqrt{2\beta \cdot x \cdot (1 - x)} + 4\beta\big\}$
\end{itemize}}
\end{minipage}}

\begin{proof}
For convenience, we only reason about the function $\eshade(\cdot)$. The same arguments can be applied to the other function $\shade(\cdot)$ as well. It suffices to consider the case that $\eshade(x) < 1$, in which we must have $x + \sqrt{2\beta \cdot x \cdot (1 - x)} < 1$ and thus
\begin{align*}
\mbox{$\sqrt{2\beta} < \frac{1 - x}{x \cdot (1 - x)} = \sqrt{x^{-1} - 1}$}
\end{align*}
Take the derivative of the function $\eshade$:
\begin{align*}
\mbox{$\frac{\d}{\d x} \eshade(x)$}
& = \mbox{$1 + \frac{1}{2} \cdot \sqrt{2\beta} \cdot \big(\sqrt{x^{-1} - 1} - \frac{1}{\sqrt{x^{-1} - 1}}\big)$} \\
& \geq \mbox{$1 + \frac{1}{2} \cdot \sqrt{2\beta} \cdot \big.\big(z - \frac{1}{z}\big)\big|_{z = \sqrt{2\beta}}$}
= \mbox{$\beta + \frac{1}{2}$}
> 0,
\end{align*}
where the first inequality is due to the above lower bound for $\sqrt{x^{-1} - 1}$ and because $y = z - \frac{1}{z}$ is an increasing function when $z > 0$. This completes the proof of \Cref{fact:shade_function}.
\end{proof}

\subsection{Proof of Lemma~\ref{lem:empirical}}
\label{subapp:lem:empirical}

Lemma~\ref{lem:empirical} is enabled by Bernstein's inequality \cite{B24}, which is given in \Cref{fact:bernstein_ineq}.

\vspace{.1in}
\noindent\fbox{
\begin{minipage}{0.977\textwidth}
\begin{fact}[Bernstein's Inequality]
\label{fact:bernstein_ineq}
Given i.i.d.\ random variables $X_1, X_2, \cdots, X_t, \cdots, X_m$ such that $\big|X_t - \E[X_t]\big| \leq \mathrm{M}$ for some constant $\mathrm{M} \in \RRP$, let $\overline{X} \eqdef \frac{1}{m} \cdot \sum_{t = 1}^m X_t$, then:
\begin{align*}
& \mbox{$\Pr\big[\big|\overline{X} - \E[X_t]\big| > \sigma\big] \leq 2 \cdot \exp\big(-\frac{m \cdot \sigma^2}{2 \cdot \Var[X_t] + \frac{2}{3} \cdot \mathrm{M} \cdot \sigma}\big)$},
&& \forall \sigma \in \RRP.
\end{align*}
\end{fact}
\end{minipage}}

\vspace{.1in}
\noindent\fbox{
\begin{minipage}{0.977\textwidth}
{\bf Lemma~\ref{lem:empirical}.}
{\em With $(1 - \delta)$ confidence, for both $i \in \{1,~2\}$, the following holds for the $i$-th highest CDF $\jdistri$ and its empirical counterpart $\edistri$: for any value $\val \in \RRP$,}
\begin{align*}
\mbox{$\big|\edistri(\val) - \jdistri(\val)\big|
\leq \sqrt{2\beta \cdot \jdistri(\val) \cdot \big(1 - \jdistri(\val)\big)} + \beta$}.
\end{align*}
\end{minipage}}

\begin{proof}
Let $\bsamplet[0] \eqdef 0$ and $\bsamplet[m + 1] \eqdef \infty$. Denote by $\bsamplet[1] \leq \bsamplet[2] \leq \cdots \leq \bsamplet[m]$ an re-ordering of the $i$-th highest sample $\hsamplei = (\hsampleti{t}{i})_{t \in [m]}$. Based on these, we can partition all the non-negative values $\val \in \RRP$ into $(m + 1)$ segments\footnote{A segment $[\bsamplet,~\bsamplet[t + 1])$ would be empty when $\bsamplet = \bsamplet[t + 1]$. Even so, the proof still works.}, namely $[\bsamplet,~\bsamplet[t + 1])$ for all $t \in [0:~m]$. Of course, every partition value $\bval = \bsamplet$ (that $t \notin \{0,~m + 1\}$) presents exactly one sample entry $\hsampleti{t}{i}$, at which the empirical $i$-th highest CDF $\edistri$ has a probability mass of $\frac{1}{m}$. Thus, for any segment $[\bsamplet,~\bsamplet[t + 1])$ and any value $\val \in \RRP$ belonging to it, we have
\begin{align}
\label{eq:lem:empirical:1}
\mbox{$0 \leq \edistri(\val) - \edistri(\bsamplet) \leq \edistri(\bsamplet[t + 1]) - \edistri(\bsamplet) \leq \frac{1}{m} \leq \frac{\ln(8m / \delta)}{3m} = \beta / 3$},
\end{align}
where the last inequality holds whenever the sample complexity $m \geq 3 \geq \frac{e^3}{8} \geq \frac{e^3}{8} \cdot \delta$.

Actually, for every partition value $\bval = \bsamplet$ that $t \in [0:~m + 1]$, we can establish a stronger concentration inequality: let $a \eqdef \sqrt{\jdistri(\bval) \cdot \big(1 - \jdistri(\bval)\big)}$ and $b \eqdef \sqrt{2\beta} \cdot a + \frac{2\beta}{3}$, then
\begin{align}
\label{eq:lem:empirical:2}
\mbox{$\Pr\big[\big|\edistri(\bval) - \jdistri(\bval)\big| > b\big] \leq \frac{\delta}{4m} \leq \frac{\delta}{2 \cdot (m + 1)}$}.
\end{align}
To see so, let us probe the i.i.d.\ Bernoulli random variable $X_t \eqdef \indicator(\hsampleti{t}{i} \leq \bval)$ for each $t \in [m]$. One can easily check that $\frac{1}{m} \cdot \sum_{t = 1}^m X_t = \edistri(\bval)$ and $\E[X_t] = \jdistri(\bval)$. Hence, we invoke Bernstein's inequality, with the parameters $\sigma = m \cdot b$, $\Var[X_t] = a^2$ and $\mathrm{M} = 1$:
\begin{align*}
\mbox{$\Pr\big[\big|\edistri(\bval) - \jdistri(\bval)\big| > b\big] \leq 2 \cdot \exp\big(-\frac{m \cdot b^2}{2 \cdot a^2 + \frac{2}{3} \cdot b}\big)$}.
\end{align*}
Since $\beta = \frac{\ln(8m / \delta)}{m}$ and $b = \sqrt{2\beta} \cdot a + \frac{2\beta}{3}$, we infer inequality~\eqref{eq:lem:empirical:2} as follows:
\begin{align*}
& \mbox{$2 \cdot \exp\big(-\frac{m \cdot b^2}{2 \cdot a^2 + \frac{2}{3} \cdot b}\big) \leq \frac{\delta}{4m}$}
&& \Leftrightarrow
&& \mbox{$b^2 - \frac{2\beta}{3} \cdot b - 2\beta \cdot a^2 \geq 0$} \\
& && \Leftarrow
&& \mbox{$\lhs = \frac{a}{3} \cdot (8\beta)^{3 / 2} \geq 0$},
\end{align*}
Due to inequality~\eqref{eq:lem:empirical:2} and the union bound (over all partition value $\bval = \bsamplet$ that $t \in [0:~m + 1]$), with at least $(1 - \delta / 2)$ confidence, we have
\begin{align}
\label{eq:lem:empirical:3}
& \big|\edistri(\bsamplet) - \jdistri(\bsamplet)\big| \leq b,
&& \forall t \in [0:~m + 1].
\end{align}
From now on, we safely assume inequality~\eqref{eq:lem:empirical:3}. In that any value $\val \in \RRP$ belongs to a unique segment $[\bsamplet,~\bsamplet[t + 1])$ that $t \in [0:~m]$,
\begin{align*}
& \mbox{$\jdistri(\val)
\geq \jdistri(\bsamplet)
\overset{\eqref{eq:lem:empirical:3}}{\geq} \edistri(\bsamplet) - b
\overset{\eqref{eq:lem:empirical:1}}{\geq} \edistri(\val) - (b + \beta / 3)$} \\
& \mbox{$\jdistri(\val)
\leq \jdistri(\bsamplet[t + 1])
\overset{\eqref{eq:lem:empirical:3}}{\leq} \edistri(\bsamplet[t + 1]) + b
\overset{\eqref{eq:lem:empirical:1}}{\leq} \edistri(\val) + (b + \beta / 3)$}
\end{align*}
That is, $|\edistri(\val) - \jdistri(\val)| \leq b + \beta / 3 = \sqrt{2\beta} \cdot a + \beta$. With at least $(1 - \delta)$ confidence, this inequality holds for both $i \in \{1,~2\}$ (by the union bound). This completes Lemma~\ref{lem:empirical}.
\end{proof}

\subsection{Proof of Lemma~\ref{lem:dominate:distr}}
\label{subapp:lem:dominate:distr}

\noindent\fbox{
\begin{minipage}{0.977\textwidth}
{\bf Lemma~\ref{lem:dominate:distr}.}
{\em In the case of Lemma~\ref{lem:empirical}, which happens with $(1 - \delta)$ confidence, for both $i \in \{1,~2\}$, the following holds for the empirical $i$-th highest CDF $\sedistri$:
\begin{enumerate}[leftmargin = 1.3em, font = {\em \bfseries}]
\item $\sedistri(\val) \geq \jdistri(\val)$ for all value $\val \in \RRP$, i.e., $\sedistri$ is dominated by the given $i$-th highest CDF $\jdistri$.
\item $\sedistri(\val) \leq \sdistri(\val)$ for all value $\val \in \RRP$, i.e., $\sedistri$ dominates the shaded $i$-th highest CDF $\sdistri$.
\end{enumerate}}
\end{minipage}}

\begin{proof}
Note that $\beta = \frac{\ln(8m / \delta)}{m} \ll 1$. For brevity, here we write $\edistri = \edistri(\val)$, $\sedistri = \sedistri(\val)$, $\jdistri = \jdistri(\val)$ and $\sdistri = \sdistri(\val)$. We justify Lemma~\ref{lem:dominate:distr} conditioned on the inequality in Lemma~\ref{lem:empirical}:
\begin{align*}
\mbox{$\big|\edistr - \jdistr\big|
\leq \sqrt{2\beta \cdot \jdistr \cdot (1 - \jdistr)} + \beta$}.
\end{align*}
{\bf Item~1: $\sedistri = \eshade(\edistri) \geq \jdistri$ for all $\val \in \RRP$.}
Since $\eshade(\cdot)$ is a non-decreasing function (Part~1 of \Cref{fact:shade_function}), it suffices to handle the case that the inequality in Lemma~\ref{lem:empirical} is an equality:
\begin{align}
\label{eq:lem:dominate:distr:edistr:0}
\edistri = \max \big\{0,~~\jdistri - \sqrt{2\beta \cdot \jdistri \cdot (1 - \jdistri)} - \beta\big\}.
\end{align}
	
\noindent
{\bf When $\edistri = 0$.}
We have $\sedistri = \eshade(\edistri) = \eshade(0) = 4\beta$ and $\jdistri - \sqrt{2\beta \cdot \jdistri \cdot (1 - \jdistri)} - \beta \overset{\eqref{eq:lem:dominate:distr:edistr:0}}{\leq} 0$, or equivalently,
\begin{align}
\label{eq:lem:dominate:distr:edistr:1}
\sqrt{2\beta \cdot (1 / \jdistri - 1)} + \beta / \jdistri \geq 1.
\end{align}
Obviously, the $\lhs$ of inequality~\eqref{eq:lem:dominate:distr:edistr:1} is a decreasing function in $\jdistri \in [0, 1]$. On the opposite of the lemma, assuming that $\jdistri > \sedistri = 4\beta$, we then have
\begin{align*}
\lhs \mbox{ of } \eqref{eq:lem:dominate:distr:edistr:1}
\leq \sqrt{2\beta \cdot \big[(4\beta)^{-1} - 1\big]} + \beta \cdot (4\beta)^{-1} < \sqrt{1 / 2} + 1 / 4 \approx 0.9571 < 1,
\end{align*}
which contradicts inequality~\eqref{eq:lem:dominate:distr:edistr:1}. Accordingly, $\sedistri \geq \jdistri$ when $\edistri = 0$.

\vskip .1in
\noindent
{\bf When $\edistri > 0$.}
We have $\edistri \overset{\eqref{eq:lem:dominate:distr:edistr:0}}{=} \jdistri - \sqrt{2\beta \cdot \jdistri \cdot (1 - \jdistri)} - \beta$. Rearranging this inequality leads to $2\beta \cdot \jdistri \cdot (1 - \jdistri) = (\jdistri - \edistri - \beta)^2$, or equivalently,
\[
(1 + 2\beta) \cdot \jdistri^2 - 2 \cdot (\edistri + 2\beta) \cdot \jdistri + (\edistri + \beta)^2 = 0.
\]
By solving this quadratic equation (i.e., select the larger one between the two roots) and taking into account the fact that $\jdistri \leq 1$, we have
\begin{align*}
\jdistri
& \leq \mbox{$\min \big\{1,~~\uwavered{(1 + 2\beta)^{-1}} \cdot \big[\edistri + 2\beta + \sqrt{(\edistri + 2\beta)^2 - (1 + 2\beta) \cdot (\edistri + \beta)^2}\big]\big\}$} \\
& \hspace{4.25cm} \mbox{\tt (drop $(1 + 2\beta)^{-1}$ term and rearrange other terms)} \\
& \leq \mbox{$\min \big\{1,~~\edistri + 2\beta + \sqrt{2\beta \cdot \edistri \cdot (1 - \edistri) + 3\beta^2 - (2\beta^3 + 4\beta^2 \cdot \edistri)}\big\}$} \\
& \hspace{4.25cm} \mbox{\tt (since $z_1 + z_2 - z_3 \leq (\sqrt{z_1} + \sqrt{z_2})^2$ when $z_i \geq 0$)} \\
& \leq \mbox{$\min \big\{1,~~\edistri + \uwave{2\beta} + \sqrt{2\beta \cdot \edistri \cdot (1 - \edistri)} + \uwave{\sqrt{3\beta^2}}\big\}$} \\
& \hspace{4.25cm} \mbox{\tt (since $2 + \sqrt{3} < 4$)} \\
& \leq \mbox{$\min \big\{1,~~\edistri + \sqrt{2\beta \cdot \edistri \cdot (1 - \edistri)} + \uwave{4\beta}\big\}$}
= \eshade(\edistri) = \sedistri.
\end{align*}
Thus, $\sedistri \geq \jdistri$ when $\edistri > 0$. Putting both cases together completes the proof of {\bfseries Item~1}.

\vspace{.1in}
\noindent
{\bf Item~2: $\sedistri = \eshade(\edistri) \leq \shade(\jdistri) = \sdistri$ for all $v \in \RRP$.} Recall \Cref{fact:shade_function}: $\eshade(\cdot)$ is a non-decreasing function. It suffices to deal with the case that the inequality in Lemma~\ref{lem:empirical} is an equality, namely
\begin{align}
\label{eq:lem:dominate:distr:edistr:2}
\edistri = \min \big\{1,~~\jdistri + \sqrt{2\beta \cdot \jdistri \cdot (1 - \jdistri)} + \beta\big\}.
\end{align}

\noindent
{\bf When $\edistri = 1$.} It follows that $\jdistri + \sqrt{2\beta \cdot \jdistri \cdot (1 - \jdistri)} + \beta \overset{\eqref{eq:lem:dominate:distr:edistr:2}}{\geq} 1$ and thus
\begin{align*}
\sdistri = \shade(\jdistri)
& = \min \big\{1,~~\jdistri + \sqrt{8\beta \cdot \jdistri \cdot (1 - \jdistri)} + 7\beta\big\} \\
& \geq \min \big\{1,~~\jdistri + \sqrt{2\beta \cdot \jdistri \cdot (1 - \jdistri)} + \beta\big\}
= 1 \geq \sedistri.
\end{align*}
	
\noindent
{\bf When $\edistri < 1$.} It follows that $\edistri \overset{\eqref{eq:lem:dominate:distr:edistr:2}}{=} \jdistri + \sqrt{2\beta \cdot \jdistri \cdot (1 - \jdistri)} + \beta$ and thus
\begin{align}
\notag
\edistri \cdot (1 - \edistri)
& = \jdistri \cdot (1 - \jdistri) + \big[\sqrt{2\beta \cdot \jdistri \cdot (1 - \jdistri)} + \beta\big] \cdot (1 - 2 \cdot \jdistri) \\
\notag
& \hspace{2.55cm} - \big[\sqrt{2\beta \cdot \jdistri \cdot (1 - \jdistri)} + \beta\big]^2 \\
\notag
& \leq \jdistri \cdot (1 - \jdistri) + \sqrt{2\beta \cdot \jdistri \cdot (1 - \jdistri)} + \beta \\
\label{eq:lem:dominate:distr:edistr:4}
& \leq \big[\sqrt{\jdistri \cdot (1 - \jdistri)} + \sqrt{\beta}\big]^2.
\end{align}
Combining everything together leads to
\begin{align}
\notag
\edistri + \sqrt{2\beta \cdot \edistri \cdot (1 - \edistri)} + 4\beta
& \overset{\eqref{eq:lem:dominate:distr:edistr:4}}{\leq} \edistri + \sqrt{2\beta \cdot \jdistri \cdot (1 - \jdistri)} + (\sqrt{2} + 4) \cdot \beta \\
\notag
& \hspace{3.45pt}=\hspace{3.45pt} \jdistri + 2 \cdot \sqrt{2\beta \cdot \jdistri \cdot (1 - \jdistri)} + (\sqrt{2} + 5) \cdot \beta \\
\label{eq:lem:dominate:distr:edistr:5}
& \hspace{3.45pt}\leq\hspace{3.45pt} \jdistri + \sqrt{8\beta \cdot \jdistri \cdot (1 - \jdistri)} + 7\beta.
\end{align}
So, $\sedistri = \min\{1,~\lhs \mbox{ of } \eqref{eq:lem:dominate:distr:edistr:5}\} \leq \min\{1,~\rhs \mbox{ of } \eqref{eq:lem:dominate:distr:edistr:5}\} = \sdistri$. We thus conclude {\bfseries Item~2}.
\end{proof}

\section{Continuous \texorpdfstring{$\lambda$-Regular}{} Setting}
\label{app:extension}

This appendix sketches out how to establish the $\cO\big(\eps^{-2} \cdot (\ln\eps^{-1} + \ln\delta^{-1})\big)$ sample complexity upper bound of \cref{alg:main} in the {\em $\lambda$-regular} setting. By standard notion, for a certain $0 < \lambda < 1$, a continuous distribution $\distri[j]$ is $\lambda$-regular when $\mathrm{H}_{j}(\val) \eqdef \big(1 - \distri[j](\val)\big)^{-\lambda}$ is a convex function on its support $\val \in \supp(\distri[j])$. Similar to the regular case and the $\mhr$ case, we assume the distributions $\jdistr = \{\distri\}_{j \in [n]}$ to be independent, and scale the instance so that $\max_{\val \in \RRP} \big\{\val \cdot \big(1 - \jdistri[1](\val)\big)\big\} = 1$. Thus, Lemma~\ref{lem:first_second} and \Cref{fact:regular:ar,fact:regular:1,fact:regular:2} still holds. From the $\lambda$-regularity, we can derive the next Lemma~\ref{lem:alpha_strong}.

\begin{lemma}
\label{lem:alpha_strong}
The following holds for any continuous $\lambda$-regular instance $\jdistr = \{\distri\}_{j \in [n]}$: given any value $u > 1$, the highest CDF $\jdistri[1](\val) \geq 1 - (\frac{u}{\val})^{1 / \lambda} \cdot \ln(\frac{u}{u - 1})$ for any larger value $\val \geq u$.
\end{lemma}

That is, the highest CDF $\jdistri[1]$ must have a {\em superlinearly-decay} tail $(\frac{u}{\val})^{1 / \lambda} \cdot \ln(\frac{u}{u - 1}) = \cO(\val^{-1 / \lambda})$. Although heavier than the {\em exponentially-decay} tail in the $\mhr$ case, this still suffices to prove a sample complexity upper bound of $\cO\big(\mathcal{C}_\lambda \cdot \eps^{-2} \cdot (\ln\eps^{-1} + \ln\delta^{-1})\big)$, where $\mathcal{C}_\lambda$ is some absolute constant given by $\lambda \in (0, 1)$.

To achieve so, we shall select a careful anchoring point $u > 1$ in Lemma~\ref{lem:alpha_strong} and adopt the almost same arguments as in the $\mhr$ case (recall Lemmas~\ref{lem:mhr:3} to \ref{lem:mhr:2}). When $\lambda \to 1$, the built-in constant $\mathcal{C}_{\lambda}$ can be arbitrarily large -- since the $\lambda$-regular family finally expands to the regular family, Part~3 of \Cref{thm:upper} declares a truly greater sample complexity of $\tTheta(\eps^{-3})$.

\begin{figure}[htbp]
\centering
\begin{tikzpicture}[thick, smooth, scale = 2.75]
\draw[->] (0, 0) -- (2.1, 0);
\draw[->] (0, 0) -- (0, 1.1);
\node[above] at (0, 1.1) {\small $\mathrm{y}$};
\node[right] at (2.1, 0) {\small $\val$};
\node[left] at (0, 0) {\small $0$};
\node[anchor = 155, color = blue] at (2, 0.8919) {\small $\mathrm{H}_{j}(\val) = \big(1 - \distri[j](\val)\big)^{-\lambda}$};
\draw[color = blue, domain = 0: 1.7] plot (\x, {0.375 * \x^2});
\draw[color = red] (0, 0) -- (2, 1.0607);
\node[below] at (1.4142, 0) {\small $u$};
\draw[dashed] (1.4142, 0) -- (1.4142, 0.75);
\draw[dashed] (1.4142, 0.75) -- (0, 0.75);
\draw[very thick] (1.4142, 0) -- (1.4142, 1pt);
\draw[very thick] (0, 0.75) -- (1pt, 0.75);
\draw[color = black, fill = green] (0, 0) circle(0.5pt);
\draw[color = black, fill = green] (1.4142, 0.75) circle(0.5pt);
\end{tikzpicture}
\caption{Demonstration for the reduction in the proof of Lemma~\ref{lem:alpha_strong}.}
\label{fig:lem:alpha_strong}
\end{figure}
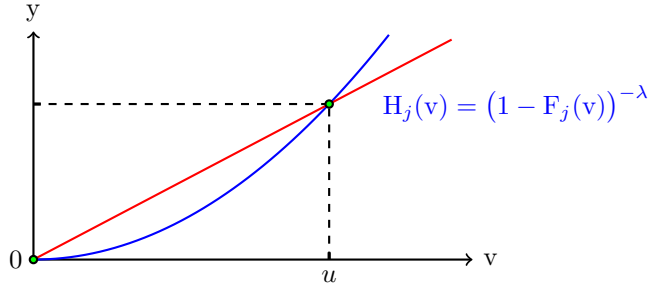

\begin{proof}
We first apply a reduction to the distributions $\jdistr = \{\distri\}_{j \in [n]}$ such that, compared to the original highest CDF $\jdistri[1]$, the resulting highest CDF decreases point-wise for any value $\val \geq u$. The reduction is illustrated in the following \Cref{fig:lem:alpha_strong}. Since $\mathrm{H}_{j}(\val) = \big(1 - \distri[j](\val)\big)^{-\lambda}$ is a convex function, define $b_j \eqdef \frac{1}{u} \cdot \mathrm{H}_j(u) > 0$, then
\begin{align}
\label{eq:alpha_strong:1}
b_j \cdot \val \leq \mathrm{H}_{j}(\val) = \big(1 - \distri[j](\val)\big)^{-\lambda}
\qquad\Rightarrow\qquad
\distri(\val) \geq 1 - (b_j \cdot \val)^{-1 / \lambda},
\end{align}
for any value $\val \geq u$, with all the equalities hold when $\val = u$. For these reasons, we have
\begin{align*}
\ln\jdistri[1](\val)
& = \mbox{$\sum_{j \in [n]} \ln\distri(\val)$}
&& \mbox{\tt (as $\jdistri[1](\val) = \prod_{j \in [n]} \distri(\val)$)} \\
& \overset{\eqref{eq:alpha_strong:1}}{\geq} \mbox{$\sum_{j \in [n]} \ln\big[1 - (b_j \cdot \val)^{-1 / \lambda}\big]$} \\
& = \mbox{$-\sum_{j \in [n]} \sum_{p = 1}^{\infty} \frac{1}{p} \cdot (b_j \cdot u)^{-p / \lambda} \cdot (\frac{\val}{u})^{-p / \lambda}$}
&& \mbox{\tt (Taylor series)} \\
& \geq \mbox{$-(\frac{\val}{u})^{-1 / \lambda} \cdot \sum_{j \in [n]} \sum_{p = 1}^{\infty} \frac{1}{p} \cdot (b_j \cdot u)^{-p / \lambda}$}
&& \mbox{\tt ($p \geq 1$ and $(\frac{\val}{u}) \geq 1$)} \\
& = \mbox{$(\frac{u}{\val})^{1 / \lambda} \cdot \ln\prod_{j \in [n]} \big[1 - (b_j \cdot \val)^{-1 / \lambda}\big]$}
&& \mbox{\tt (Taylor series)} \\
& \overset{\eqref{eq:alpha_strong:1}}{=} \mbox{$(\frac{u}{\val})^{1 / \lambda} \cdot \ln\jdistri[1](u)$}
&& \mbox{\tt (equality condition holds)} \\
& \geq \mbox{$(\frac{u}{\val})^{1 / \lambda} \cdot \ln\big(1 - \frac{1}{u}\big)$}.
&& \mbox{\tt (\Cref{fact:regular:1}: $\jdistri[1](u) \geq 1 - \frac{1}{u}$)}
\end{align*}
After being rearranged, this inequality becomes
\begin{align*}
\jdistri[1](\val)
\geq \mbox{$e^{-(\frac{u}{\val})^{1 / \lambda} \cdot \ln(\frac{u}{u - 1})}$}
\geq \mbox{$1 - (\frac{u}{\val})^{1 / \lambda} \cdot \ln(\frac{u}{u - 1})$},
\end{align*}
where the last inequality is because $e^{-z} \geq 1 - z$. This completes the proof of Lemma~\ref{lem:alpha_strong}.
\end{proof}

\end{document}